\documentclass[reprint, aps, prx, longbibliography, notitlepage]{revtex4-1}
\usepackage{amsmath,amssymb, bm,graphicx}
\usepackage{graphics}
\usepackage{float} 
\usepackage[stable]{footmisc}
\usepackage{soul}
\usepackage[colorlinks, linkcolor= blue, citecolor = blue, urlcolor=blue]{hyperref}
\usepackage{makecell}
\usepackage{physics}
\usepackage{notes2bib}
\usepackage{appendix}
\usepackage{ragged2e}
\usepackage{caption}
\usepackage{multirow} 
\usepackage{array}      
\usepackage{braket}
\usepackage{subcaption}

\def\be{\begin{equation}}
\def\ee{\end{equation}}
\def\bea{\begin{eqnarray}}
\def\eea{\end{eqnarray}}

\begin{document}

\title{Spin-split magnon bands induce pure spin current in insulating altermagnets}
\author{Sankar Sarkar}
\email{sankars24@iitk.ac.in}
\affiliation{Department of Physics, Indian Institute of Technology Kanpur, Kanpur-208016, India}
\author{Amit Agarwal}
\email{amitag@iitk.ac.in}
\affiliation{Department of Physics, Indian Institute of Technology Kanpur, Kanpur-208016, India}

\begin{abstract}
Altermagnets offer a promising platform for dissipationless spin transport by combining zero net magnetization with spontaneous non-relativistic spin-split bands. However, their magnonic transport properties remain largely unexplored. Here, we develop a quantum-kinetic theory for thermally driven magnon currents that cleanly separates Berry-curvature–driven intrinsic contributions from Drude-like scattering-dependent terms. Applying this framework to a collinear honeycomb antiferromagnet with anisotropic next-nearest-neighbor exchange and Dzyaloshinskii–Moriya interaction, we reveal spin-split magnon bands that support both intrinsic and extrinsic spin Nernst and Seebeck currents. For realistic parameters, we predict a sizable magnon spin-splitting angle (about 3.3 degrees) and a pure transverse spin current capable of exerting a strong spin-splitter torque suitable for magnetization switching.
\end{abstract}
\maketitle
\section{Introduction}

Spin-based transport in insulating magnets offers an attractive route to energy-efficient information processing and emerging quantum technologies \cite{Sarma_2004,Bader_2010,hirohata_2020}. Magnons, the bosonic spin-wave excitations of an ordered spin lattice, carry angular momentum without charge and propagate without Joule heating. This enables ultra-low-dissipation spintronic devices \cite{Maekawa_2012,blundell_2001,Baltz_2018}. Their electrical neutrality facilitates the generation and manipulation of pure spin currents in magnetic insulators \cite{Maekawa_2012,chumak_2015,Baltz_2018}. However,  conventional magnets impose intrinsic constraints. In ferromagnets, a magnon’s spin is locked to the magnetization direction, limiting flexibility \cite{Cheng_2016}. In antiferromagnets, spin transport hinges on relativistic spin–orbit coupling (SOC), which is much smaller than exchange interactions and this limits both  efficiency and tunability \cite{Cheng_2016,Vladimir_2016,Rezende_2019,Rezende_2016,Shen_2019,Xiao_2010,Rezende_2014,Uchida_2010,Seshadri_2018}. These limitations motivate the search for alternative magnetic platforms capable of robust, tunable, and  dissipationless spin transport without relying on either net magnetization or strong SOC.

A recently proposed class of materials, altermagnets (AMs), offer a promising route to overcoming the limitations of conventional magnetic systems. Like antiferromagnets, AMs have no net magnetization because of compensating magnetic sublattices, even though their band structures exhibit spontaneous spin splitting in momentum space, similar to that in ferromagnets \cite{Sinova_2022,Fender_2025,Gomonay_2024,Cui_2023,krempasky_2024,Liu_2024,Jairo_2022}. This distinctive behaviour arises from anisotropic exchange interactions that originate either from non-magnetic atoms in the local crystal environment or from orbital anisotropy of the magnetic ions \cite{Cui_2023,Brekke_2023,Alberto_2024,Lee_2024,Camerano_2025, camerano_2025_2}. The magnonic bands in altermagnets host two distinct magnon modes with opposite spin angular momentum, making AMs a natural platform for encoding, transporting, and manipulating spin information \cite{Cheng_2016,Vladimir_2016}.

Here, we study thermally driven magnonic spin transport using a density-matrix–based quantum-kinetic framework \cite{Harsh_2023, Harsh_2023_2, Harsh_2024, Sayan_2025, Sayan_2024}. We derive analytic expressions for the magnon spin Nernst and spin Seebeck conductivities, showing that the resulting spin currents decompose into an \emph{intrinsic} (Berry-curvature–driven) and an \emph{extrinsic} (Drude-like, scattering-dependent) part. To illustrate the physics, we analyze a collinear altermagnet on a honeycomb lattice with anisotropic next-nearest-neighbor (NNN) exchange and a Dzyaloshinskii–Moriya interaction (DMI). We find that the extrinsic response originates solely from the \emph{non-relativistic} anisotropic exchange couplings, while the intrinsic contribution requires the  \emph{relativistic} DMI to be finite. We demonstrate the thermal gradient driven magnonic spin-splitter effect that generates a \emph{pure} transverse spin current with a spin-splitting angle of around $ 3.3^{\circ}$. This current can exert a sizeable spin-splitter torque (SST) on an adjacent ferromagnetic layer. We estimate the effective magnetic field generated by the SST to be around $ 2$ mT, sufficient for enabling thermally driven magnetization switching. These results position magnons as a robust platform for ultra-low-dissipation spin transport in insulating altermagnets and a concrete route toward next-generation spin-caloritronic devices.

\section{Thermally induced magnonic Spin Current}\label{Magnon Spin Current from Thermal Gradient}

Here, we study the spin current \cite{Maekawa_2012} generated by spin-split magnon bands in altermagnets \cite{Jairo_2022}. The magnon spin currents dominate the spin responses in magnetic insulators \cite{Maekawa_2012,chumak_2015,Rezende_2014,Rezende_2016,Uchida_2010,Rezende_2019,Xiao_2010}. We calculate the magnon spin current induced by a temperature gradient using the density matrix-based quantum kinetic theory framework \cite{Harsh_2023,Harsh_2023_2}. 

In thermal equilibrium and in the absence of any perturbations, the occupancy of the magnon bands obeys the Bose-Einstein (BE) distribution, 
\(
f_n(\mathbf{k}) = \left[ e^{\varepsilon_n(\mathbf{k})/k_B T} - 1 \right]^{-1}.
\) Here, \( \varepsilon_n(\mathbf{k}) \) is magnon band energy for the $n$th band at Bloch momentum $\mathbf{k}$, \( k_B \) is Boltzmann constant, and \( T \) is the absolute temperature. In the BE distribution, the chemical potential \( \mu \) is set to zero, reflecting the non-conserving nature of magnon number. A perturbing temperature gradient drives the system out of equilibrium, leading to a time-dependent magnon density matrix \( \rho(\mathbf{k}, t) \). The time evolution of \( \rho(\mathbf{k}, t) \) is governed by the quantum kinetic equation~\cite{Harsh_2023,Harsh_2023_2,Sekine_2020,Culcer_2017,boyd_2008}, 
\begin{equation}
    \frac{\partial \rho(\mathbf{k},t)}{\partial t} + \frac{i}{\hbar} [\mathcal{H},~ \rho] + \frac{\rho(\mathbf{k},t)}{\tau} = D_T[\rho(\mathbf{k})]~.
    \label{quantum-kinetic-equation1}
\end{equation}
In the third term of Eq.~(\ref{quantum-kinetic-equation1}), we have used the relaxation-time approximation for incorporating magnon scattering. Additionally, we have assumed a momentum-independent relaxation time \( \tau \) for simplicity. A momentum or energy dependence of the scattering timescale can be easily included in our calculations and does not change the qualitative aspects of our results. 

The right-hand side of Eq.~(\ref{quantum-kinetic-equation1}) captures the influence of the temperature gradient. Following Refs.~\cite{Tatara_2015,Sekine_2020}, we incorporate the temperature gradient via a thermal vector potential \( \mathbf{A}_T \)~\cite{Tatara_2015}, which modifies the canonical momentum as \( \mathbf{p} \rightarrow \mathbf{p} + \mathcal{H} \cdot \mathbf{A}_T \). This leads to the thermal driving term, $D_T[\rho(\mathbf{k})]$~\cite{Harsh_2023,Harsh_2023_2,Sekine_2020} given by, 
\begin{equation}
    D_T[\rho (\mathbf{k})] = -\frac{1}{2\hbar}~\mathbf{E}_T \cdot \left[ \left\{ \mathcal{H}, \frac{\partial \rho}{\partial \mathbf{k}} \right\} - i \left[ \mathcal{R}, \left\{ \mathcal{H}, \rho \right\} \right] \right]~.
    \label{DT}
\end{equation}
Here, \( \mathbf{E}_T = -\partial_t \mathbf{A}_T = -\nabla T/T \) is the thermal field, and 
the operator \( \mathcal{R}_{np}(\mathbf{k}) = i \bra{u_n(\mathbf{k})} \sigma^z \nabla_{\mathbf{k}} \ket{u_p(\mathbf{k})} \) denotes the non-Abelian Berry connection for the magnon bands~\cite{Gao_2021, Sinha_2022, Debottam_2024,Kamal_2023,Harsh_2024}. The $\sigma_z$ matrix appearing in expression of Berry connection is the bosonic metric, which also appears in the eigenvalue equation discussed in the next section, Eq.~\eqref{eq:eigen_value_equation}. Throughout this work, we use the notation \( \bra{u_n(\mathbf{k})} \sigma^z \hat{\mathcal{O}} \ket{u_p(\mathbf{k})} = \mathcal{O}_{np} \) to denote matrix elements of any operator \( \hat{\mathcal{O}} \) in the magnon band basis.

To determine the nonequilibrium density matrix, we adopt a perturbative approach~\cite{boyd_2008}. We expand the non-equilibrium density matrix in powers of the thermal field strength, $    \rho(\mathbf{k}) = \rho^{(0)}(\mathbf{k}) + \rho^{(1)}(\mathbf{k}) + \cdots$. The zeroth-order term corresponds to the equilibrium distribution, and we have $ \rho^{(0)}(\mathbf{k}) = \sum_n \ket{u_n(\mathbf{k})} \bra{u_n(\mathbf{k})} f_n(\mathbf{k})$.  

The first-order correction \( \rho^{(1)}(\mathbf{k}) \propto  |E_T| \) accounts for the linear response of the system to the applied temperature gradient. Substituting this expansion in Eq.~\eqref{quantum-kinetic-equation1}, we obtain $\rho^{(1)} (\mathbf{k}, t)$ ~\cite{Harsh_2023_2,Harsh_2023,Culcer_2017,boyd_2008}. We present the details of the calculations of the first-order density matrix in Appendix \ref{density-matrix-calculations}. For a system driven by a harmonic temperature gradient, $E_T \propto e^{i \omega t}$, the first order density matrix is given by, 
\begin{align}
    \rho^{(1)}_{np} (\mathbf{k}, t) &= -\frac{E^a_T}{2\hbar}~[\varepsilon_n ~ \partial_a f_n ~\delta_{np}~(g^{-\omega}_0  e^{i\omega t}+ g^{\omega}_0 ~e^{-i\omega t})\notag \\ &\quad  + i ~ \mathcal{R}^a_{np} ~\xi_{np}~(g^{-\omega}_{np} ~e^{i\omega t} + g^{\omega}_{np} ~ e^{-i\omega t}) ~].
\end{align}
Here, we use the Einstein summation convention over the index $a$. We have defined $g^{\omega}_{np} = [1/\tau - i( \omega - \omega_{np})]^{-1}$ and $g^{\omega}_0 = [1/\tau - i\omega]^{-1}$, \( \xi_{np} = (\varepsilon_n f_n - \varepsilon_p f_p) \), $(\partial/\partial k_a) \equiv \partial_a$, and \( \omega_{np} = (\varepsilon_n - \varepsilon_p)/\hbar \).

\subsection{Magnon currents and response tensors}
Magnons carry a quantized unit of spin angular momentum, specified by $\hbar$~\cite{chumak_2015}. Consequently, the spin current is simply the magnon current (particle) multiplied by $\hbar$~\cite{Rezende_2019,Seki_2015,Shen_2019,Baltz_2018,Vladimir_2016}. We show below that the magnon current has two distinct contributions: $(i)$ an extrinsic contribution proportional to the scattering timescale ($\tau$)~\cite{Harsh_2023}, and $(ii)$ an intrinsic scattering time-independent contribution arising from the band geometric properties. 
The measurable magnon current denisty 
is given by \cite{Matsumoto_2011,Matsumoto_2011_prl,Cheng_2016,Vladimir_2016,Rezende_2019,Adachi_2013,Rezende_2014,Harsh_2023, Harsh_2023_2, Harsh_2024}
\begin{equation}
    J^{a} = \sum_{\mathbf{k}}{\rm Tr}[ v^a(\mathbf{k})~ \rho^{(1)} ] + \sum_{\mathbf{k}} \text{Tr}[ \mathbf{E}_T \times \mathbf{M_{\Omega}}]_a ~.
    \label{def_of_magnon_current}
\end{equation}
Here, $a ~ (=x,y,z)$ is the direction of magnon current and $v^a(\mathbf{k})$ is the band velocity along $a$, and Tr stands for trace. The second term in Eq.~(\ref{def_of_magnon_current}) subtracts the orbital magnetization induced local circulating current, which does not reflect in the measured current. The local circulating magnetization current contribution is induced by the Berry curvature in Eq.~(\ref{def_of_magnon_current}) \cite{Harsh_2023,Harsh_2023_2, Harsh_2024} and its explicit form for a specific band is given by \cite{Matsumoto_2011,Matsumoto_2011_prl},
\begin{equation}
    \mathbf{M}_{\mathbf{\Omega}_n} = \frac{1}{\hbar} \int_{\varepsilon_{n\mathbf{k}}}^{\infty} \varepsilon~ \frac{\partial f(\varepsilon)}{\partial \varepsilon} ~ \mathbf{\Omega}_{n} (\mathbf{k}) \, d\varepsilon~.
    \label{eq:magnetization}
\end{equation}
Here, $\mathbf{\Omega}_n (\mathbf{k}) = i~\bra{\partial_\mathbf{k} u_n} \sigma^z \ket{\partial_\mathbf{k} u_n}$ is the Berry curvature for the $n$th magnon band.

Inserting the first order density matrix, \( \rho_{np}^{(1)} (\mathbf{k}, t) \)  into Eq.~(\ref{def_of_magnon_current}), we obtain, 
\begin{align}
    J^{a}(\omega, t) &=  \sum_{\mathbf{k},n,p}~v^a_{pn}~\rho^{(1)}_{np}(\mathbf{k}) +  \sum_{\mathbf{k},n} \epsilon_{abc} ~M^c_{\mathbf{\Omega}_n} E_T^b  \notag\\
    &= \frac{\nabla_b T}{T}\times[ \sigma^{ab}(\omega)~e^{i\omega t} + \text{c.c.} ]~.\label{response-tensor_definition}
\end{align}
Here, \( \sigma^{ab} (\omega) \) denotes magnon current conductivity along \( a -\)direction when an thermal field is applied along $b-$direction. Moreover, $ \epsilon_{abc}$ denotes the anti-symmetric Levi-Civita tensor, `c.c.' denotes the complex conjugate, and the interband velocity matrix element is given by $ v^a_{pn} = v^a_{nn} \delta_{pn} + i\mathcal{R}^a_{pn}\omega_{pn} $ \cite{Harsh_2024}. The magnon current conductivity tensor, $\sigma^{ab}(\omega)$ is given by, 
\begin{align}
    \sigma^{ab} (\omega)& = \frac{1}{2 \hbar}\sum_{\mathbf{k},n} \left[~\varepsilon_n~ v^a_{nn}~ \partial_b f_n ~ g^{-\omega}_0 -2 ~ \hbar ~ \epsilon_{abc} ~ M^c_{\mathbf{\Omega}_n} ~\right] \notag \\
    & +~\frac{1}{2 \hbar}~\sum_{\mathbf{k},n,p}^{n\neq p} ~ \left[ \mathcal{G}^{ab}_{pn} - \frac{i}{2}~\Omega^{ab}_{pn} \right]~\omega_{np}~\xi_{np}~g^{-\omega}_{np}.
    \label{magnon-current-conductivity}
\end{align}
In Eq.~(\ref{magnon-current-conductivity}), we have used the \textit{quantum geometric tensor} $ Q^{ab}_{pn} = \mathcal{R}^a_{pn}\mathcal{R}^b_{np} =  \mathcal{G}^{ab}_{pn}-(i/2)\Omega^{ab}_{pn} $. Its real part  $\mathcal{G}^{ab}_{pn}$ is the band-resolved \textit{quantum metric tensor} and the imaginary part $\Omega^{ab}_{pn}$ is the band-resolved \textit{Berry curvature}~\cite{Debottam_2024,Harsh_2023,Harsh_2024,Harsh_2023_2,Kamal_2023,Pankaj_2023,Pankaj_2022,Maneesh_2024}. 

\subsection{Intrinsic and extrinsic spin current conductivity}
In this subsection, we use the previously developed magnon current conductivity ($\sigma^{ab}$) to analyze the intrinsic and extrinsic contributions to the magnon conductivity and define the magnon spin current conductivity tensor. Consider a $d.c.$ thermal field i.e., $\omega = 0$, which leads to $g^{-\omega}_0 = \tau$. Additionally, we adopt the dilute impurity approximation, where $\tau \omega_{np} \gg 1$~\cite{Harsh_2023,Harsh_2023_2}. In this regime, we can approximate  
$g^{-\omega}_{np} \approx \frac{1}{i\omega_{np}}$. Using the symmetry properties of the band-geometric quantities, 
$\mathcal{G}^{ab}_{np} = \mathcal{G}^{ab}_{pn}$, $\Omega^{ab}_{np} = - \Omega^{ab}_{pn}$ and \( \Omega_n = \sum_{p \neq n} \Omega_{np} \), in Eq.~(\ref{magnon-current-conductivity}), we obtain,
\begin{align}
    \sigma^{ab}_{\text{ext}} &= \frac{~\tau}{2}~\sum_{\mathbf{k}, n}~\left[ \varepsilon_n~v^a_{nn} ~ v^b_{nn}~\frac{\partial f_n}{\partial \varepsilon_n} \right]~,\notag\\
    \sigma^{ab}_{\text{int}} &= \frac{1}{2 \hbar}~\sum_{\mathbf{k},n}~\epsilon_{abc}~\left[ \Omega^{c}_{n}~\varepsilon_n ~f_n  -2~\hbar~  M^c_{\mathbf{\Omega}_n}~\right]~. \label{extrinsic-and-intrinsic}
\end{align}
The extrinsic contribution to the magnon current conductivity $\sigma^{ab}_{\text{ext}} \propto \tau$, it depends solely on the band energies and velocities, and is a Fermi surface phenomenon. In contrast, the intrinsic contribution $\sigma^{ab}_{\text{int}} \propto \tau^0$ arises from band geometric quantities and involves Fermi sea contributions. The presence of the Levi-Civita tensor in $\sigma^{ab}_{\text{int}}$ ensures that it always yields a transverse (Nernst-type) response. In contrast, $\sigma^{ab}_{\text{ext}}$ may contribute to both longitudinal (Seebeck) and transverse (Nernst) components depending on the band structure and symmetry. 

Having established the magnon current from scattering and band geometry contributions, 
we now define the corresponding magnon spin current conductivity. For the spin current with spin polarization along $z-$axis, the conductivity is given by,
\begin{align}
    \sigma^{z;ab}_{\text{ext}} &= -\frac{\hbar~\tau}{2}~\sum_{\mathbf{k}, n}~\left[ \varepsilon_n~v^a_{nn} ~ v^b_{nn}~s_n~\frac{\partial f_n}{\partial \varepsilon_n} \right]~,\notag\\
    \sigma^{z;ab}_{\text{int}} &= -\frac{1}{2}~\sum_{\mathbf{k},n}~\epsilon_{abc}~s_n~\left[ \Omega^{c}_{n}~\varepsilon_n ~f_n  -2~\hbar~  M^c_{\mathbf{\Omega}_n}~\right]~. \label{extrinsic-and-intrinsic_spin_conductivity}
\end{align}
Here, \(s_{n} = \pm 1\) is the spin polarization of the $n$th band along the $z$-axis.

Having developed the framework for calculating the magnon spin conductivity, we now demonstrate this in an insulating altermagnet. 


\section{Generalized Spin Model for Altermagnetic Magnons}
\label{sec:spin_model}
In the absence of any external magnetic field or SOC, the antiferromagnetic magnon bands are degenerate. However, in altermagnets, this degeneracy is lifted by anisotropic exchange interactions between inequivalent magnetic sites~\cite{Cui_2023, Gomonay_2024}. To study the magnon-mediated spin transport in insulating alternagnets, we propose a generalized spin model on a honeycomb lattice with collinear antiferromagnetic order. The spin Hamiltonian captures momentum-dependent spin splitting through anisotropic NNN exchange couplings and DMI.

\begin{figure*}[t]
    \centering    \includegraphics[width=\linewidth]{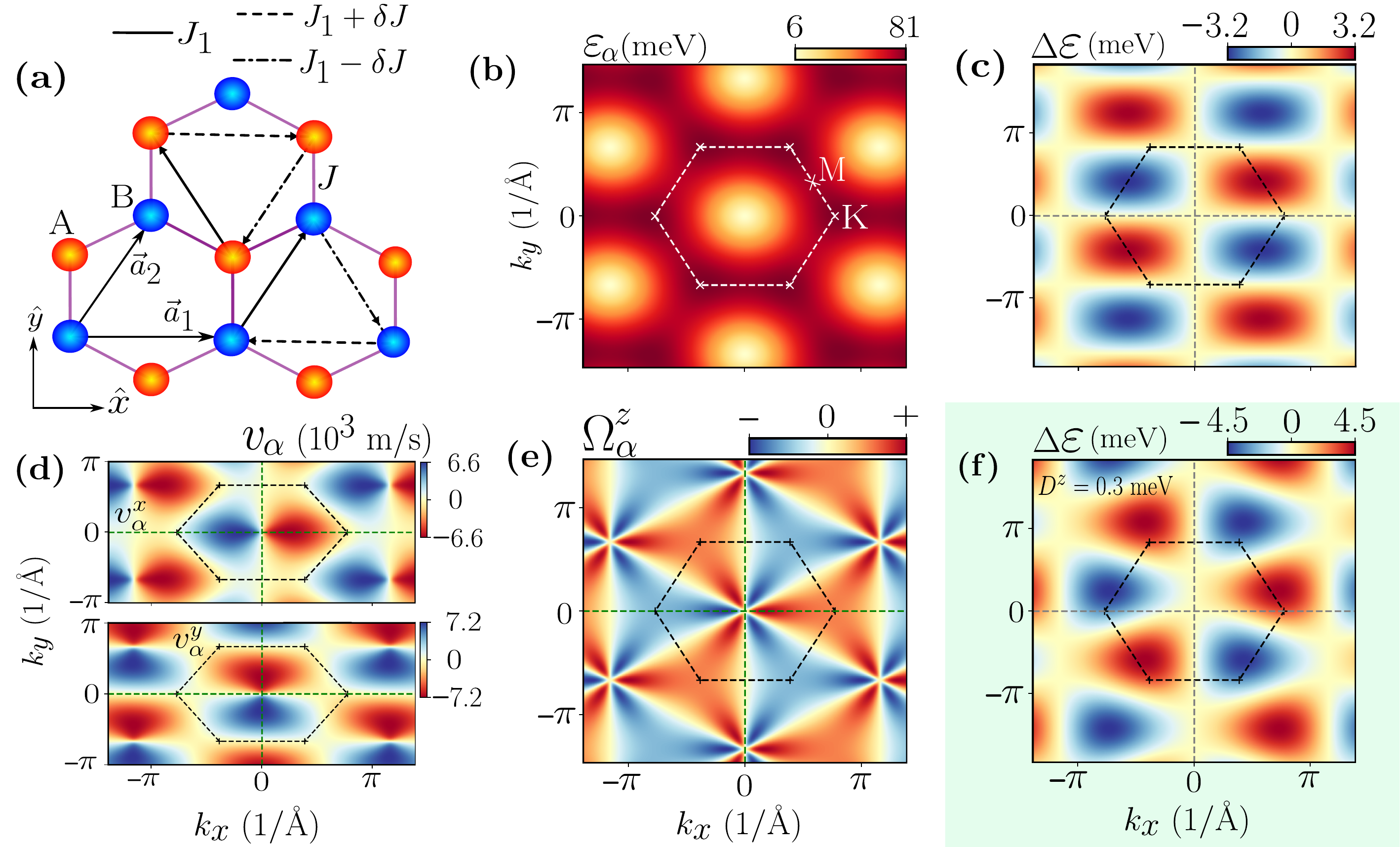}
    \caption{\justifying
    \textbf{Honeycomb AM: Magnon dispersion and band geometric properties $(S=1)$.}
\textbf{(a)} Honeycomb lattice with antiferromagnetically  aligned spins on sublattices A (spin-up) and B (spin-down). NN exchange coupling ($J > 0$) is same for all three NN bonds. $J_1, J_1-\delta J$ and $J_1 + \delta J$ denote NNN intra-sublattice exchange couplings. Real-space lattice vectors are ${\bf a}_1 = (\sqrt{3}a, 0)$, ${\bf a}_2 = a(\sqrt{3}/2, 3/2)$. \textbf{(b)} Magnon dispersion for the $\alpha$-mode over the full Brillouin zone, with dotted lines indicating the hexagonal Brillouin zone boundary. \textbf{(c)} $k$-space distribution of  magnon band splitting, $\Delta \varepsilon = \varepsilon_{\alpha} - \varepsilon_{\beta}$, for $D^z = 0$, showing a characteristic $d$-wave-like structure. \textbf{(d)} Components of the band velocity of the $\alpha$-mode along $x$- and $y$- directions. \textbf{(e)} Berry curvature for the $\alpha$-mode over the Brillouin zone. \textbf{(f)} $k$-space distribution of  magnon band splitting, $\Delta \varepsilon$ for $D^z = 0.3$. We have used the following parameters, $\{J,J_1, \delta J, K \} = \{12.4, -5.48, 0.8, -0.3\}$ meV, $D^z = 0$ for \textbf{(b - e)}, $D^z = 0.3$ for \textbf{(f)} and lattice parameter, $a = 1$ \AA.
\label{magnon-bands}}
\end{figure*}

The spin Hamiltonian is defined as, 
\begin{align}
\mathcal{H} &= J \sum_{\langle i \in A, j \in B \rangle} \mathbf{S}_i \cdot \mathbf{S}_j + K \sum_i (S_i^z)^2 \notag \\
&\quad + \frac{1}{2} \sum_{m=1}^6 \sum_{i \in A} \left[ J_{mA} \, \mathbf{S}_i \cdot \mathbf{S}_{i + \delta_m} + \mathbf{D}_{mA} \cdot (\mathbf{S}_i \times \mathbf{S}_{i + \delta_m}) \right] \notag \\
&\quad + \frac{1}{2} \sum_{m=1}^6 \sum_{i \in B} \left[ J_{mB} \, \mathbf{S}_i \cdot \mathbf{S}_{i + \delta_m'} + \mathbf{D}_{mB} \cdot (\mathbf{S}_i \times \mathbf{S}_{i + \delta_m'}) \right].
\label{eq:hamiltonian}
\end{align}
Here, the first term describes nearest-neighbor (NN) antiferromagnetic exchange (\( J > 0 \)) between sublattices \( A \) and \( B \), and the uniaxial anisotropy term (\( K < 0 \)) ensures collinear ordering~\cite{blundell_2001}. The second term represents NNN interactions within sublattice \( A \), with anisotropic exchanges \( J_{mA} \), DMIs \( \mathbf{D}_{mA} = +D^z \hat{z} \) for clockwise oriented bonds along directions \( \delta_{1,2,3} \) and \( \mathbf{D}_{mA} = -D^z \hat{z} \) for anti-clockwise oriented bonds along directions \( \delta_{4,5,6} \). The third term describes NNN interactions within sublattice \( B \), with exchanges \( J_{mB} \), DMIs \( \mathbf{D}_{mB} = +D^z \hat{z} \) along directions \( \delta'_{1,2,3} \) and \( \mathbf{D}_{mB} = -D^z \hat{z} \) for \( \delta'_{4,5,6} \). The NNN position vectors for sublattice \( A \) are: 
\( \delta_1 = \left( -\sqrt{3}a/2, 3a/2 \right) \), 
\(\delta_2 = \left( -\sqrt{3}a/2, -3a/2 \right) \), and \(\delta_3 = \left( \sqrt{3}a, 0 \right)\), and  \(\delta_{4,5,6} = -\delta_{1,2,3}\). For sublattice B the NNN positions vectors are opposite of those for A i.e., $\delta^{\prime}_m = - \delta_m$ [See Fig.~\ref{magnon-bands}\textcolor{blue}{(a)}].

In principle, we can choose fully independent NNN exchange couplings to lift the magnon bands degeneracy and obtain spin-split magnon bands. However, such generality leads to cumbersome calculations. To simplify the analysis while preserving the essential physics, we adopt a symmetric anisotropic pattern for the exchange couplings. Specifically, for sublattice A, we set, 
$J_{1A} = J_{4A} = J_1$, $J_{2A} = J_{5A} = J_1 - \delta J$, and $J_{3A} = J_{6A} = J_1 + \delta J$. For sublattice B, we choose, 
$J_{1B} = J_{4B} = J_1 - \delta J$, $J_{2B} = J_{5B} = J_1$, and $J_{3B} = J_{6B} = J_1 + \delta J$, as illustrated in Fig.~\ref{magnon-bands}\textcolor{blue}{(a)}.  This specific choice relates the two sublattices via the composite symmetry operation \( [C_2 \,||\, M_y] \), where \( C_2 \) is a two-fold rotation in spin space, and \( M_y \) is a mirror reflection perpendicular to the \( y \)-axis \cite{Lee_2024}. 


The choice of distinct NNN exchange couplings (\(J_{mA}, J_{mB}\)) reflects the sublattice and bond-dependent anisotropy in altermagnets. These can arise from the local crystal environment or orbital anisotropy of magnetic ions~\cite{Sinova_2022,Cui_2023,Lee_2024}. In a honeycomb lattice, the six NNN bonds per site connect inequivalent magnetic sites within each sublattice.
This anisotropy breaks sublattice symmetry and induces momentum-dependent altermagnetic spin splitting of magnon bands without net magnetization~\cite{Gomonay_2024,Cui_2023}. For instance, in hexagonal MnTe, non-symmorphic symmetries lead to bond-dependent exchange~\cite{Lee_2024,Lovesey_2023}. 

The inclusion of DMI exclusively on NNN bonds, is motivated by the need to enhance altermagnetic splitting and for getting a finite Berry curvature induced intrinsic magnon current response. We omit NN DMI because it typically induces non-collinear spin textures (e.g., canting), which are incompatible with the collinear antiferromagnetic ground state~\cite{Baltz_2018,Cheng_2016}. However, NNN DMI contributes to spin splitting without disrupting collinearity, as it acts within sublattices and modulates magnon dispersion~\cite{Cui_2023}.  
%
We restricted the DMI to the \(z\)-component (\(D^z\)) as it aligns with the magnetic easy axis, and stabilizes the ground state against transverse spin fluctuations~\cite{blundell_2001}. In-plane DMI components (\(D_x, D_y\)) introduce transverse spin interactions, which can lead to spiral order or other instabilities, disrupting the collinear AFM ground state~\cite{Baltz_2018}.  

\subsection{Magnon dispersion and band-splitting}

This generalized spin model produces spin-split magnon bands through both sublattice asymmetry (\(J_{mA} \neq J_{mB}\)) and DMI. To obtain the magnon dispersion, we work in the low magnon density regime and ignore magnon-magnon interactions. We use the Holstein-Primakoff (HP) transformations to express the spin Hamiltonian in terms of bosonic operators. The HP transformations for a two-sublattice antiferromagnet are~\cite{Kubo_1952}, 
\begin{align}
S_i^+ &\approx \sqrt{2S} a_i,~~ S_i^- \approx \sqrt{2S} a_i^\dagger, ~~ S_i^z = S - a_i^\dagger a_i, ~~ [i \in A], \notag \\
S_j^+ &\approx \sqrt{2S} b_j^\dagger, ~~ S_j^- \approx \sqrt{2S} b_j, ~~ S_j^z = -S + b_j^\dagger b_j, ~~ [j \in B]. 
\end{align}
Here, $S_i^{\pm} = (S_i^x \pm iS_i^y)/2$ are rising and lowering operator for spin at $i$-th site. The operators \( a_i^\dagger (b_i^\dagger) \) creates a spin deviation on sublattice $A$ ($B$), and \( a_i (b_i) \) annihilates it, satisfying bosonic commutation relations \([a_i, a_j^\dagger] = [b_i, b_j^\dagger] = \delta_{ij}\). Applying the HP transformations to Eq.~(\ref{eq:hamiltonian}) yields a real space bosonic Hamiltonian. 
To obtain the momentum space Hamiltonian, we perform a Fourier transformation,  
\begin{align}
\begin{bmatrix} a_i \\ b_i^\dagger \end{bmatrix} = \frac{1}{\sqrt{N}} \sum_{\mathbf{k}} e^{i \mathbf{k} \cdot \mathbf{r}_i} \begin{bmatrix} a_{\mathbf{k}} \\ b_{\mathbf{k}}^\dagger \end{bmatrix}.\label{Fourier_transformation}
\end{align}
This yields the Hamiltonian, \(\mathcal{H} = \sum_{\mathbf{k}} \psi_{\mathbf{k}}^\dagger \mathcal{H}_{\mathbf{k}} \psi_{\mathbf{k}}\), with \(\psi_{\mathbf{k}}^\dagger = [a_{\mathbf{k}}^\dagger, b_{\mathbf{k}}]\). Details of the derivation this Hamiltonian are presented in Appendix~\ref{magnon-bands-derivation}. The momentum-resolved Hamiltonian is given by, 
\begin{align}
\mathcal{H}_{\mathbf{k}} = \begin{bmatrix}
J' + \gamma_A(\mathbf{k}) & J S \gamma(\mathbf{k}) \\
J  S\gamma(\mathbf{k})^* & J' + \gamma_B(\mathbf{k})
\end{bmatrix}~.
\label{eq:magnon_hamiltonian}
\end{align}
Here, \( J' = (3JS  - 2KS) \), and the structure factors are defined by, 
\begin{align}
\gamma(\mathbf{k}) &= \sum_{j=1}^3 e^{i \mathbf{k} \cdot \mathbf{d}_j} = 2 e^{i a k_y / 2} \cos\left( \frac{\sqrt{3} a k_x}{2} \right) + e^{-i a k_y}, \notag \\
\gamma_A(\mathbf{k}) &= 2S\sum_{m=1}^3 \left[ J_{mA} \{ \cos(\mathbf{k} \cdot \delta_m) -1\} - D^z \sin(\mathbf{k} \cdot \delta_m)\right], \notag \\
\gamma_B(\mathbf{k}) &=2S \sum_{m=1}^3 \left[ J_{mB} \{ \cos(\mathbf{k} \cdot \delta_m) -1\} + D^z \sin(\mathbf{k} \cdot \delta_m)\right].
\end{align}
In the structure factor $\gamma(\mathbf{k})$, we use the NN positions vectors: \(\mathbf{d}_1 = \left( \sqrt{3}a/2, a/2 \right)\), \(\mathbf{d}_2 = \left( -\sqrt{3}a/2, a/2 \right)\), and \(\mathbf{d}_3 = (0, -a)\). 

The matrix $\mathcal{H}$ is non-Hermitian due to the mixed operator basis \(a_{\mathbf{k}}^{\dagger},~ b_{\mathbf{k}}\),  and it requires a bosonic Bogoliubov transformation for diagonalization. The Bogoliubov transformation is given by, 
\begin{align}
\alpha_{\mathbf{k}} = u_{\mathbf{k}} a_{\mathbf{k}} - v_{\mathbf{k}} b_{\mathbf{k}}^\dagger, \quad \beta_{\mathbf{k}} = u_{\mathbf{k}} b_{\mathbf{k}} - v_{\mathbf{k}} a_{\mathbf{k}}^\dagger~,
\end{align}
where \(\alpha_{\mathbf{k}}~ (\alpha_{\mathbf{k}}^{\dagger}) \) and \(\beta_{\mathbf{k}}~ (\beta_{\mathbf{k}}^{\dagger})\) are magnon annihilation (creation) operators for the \(\alpha\)- and \(\beta\)-magnon modes, respectively. These operators satisfy the following bosonic commutation relations: \([\alpha_{\mathbf{k}}, \alpha_{\mathbf{k}'}^\dagger] = [\beta_{\mathbf{k}}, \beta_{\mathbf{k}'}^\dagger] = \delta_{\mathbf{k} \mathbf{k}'}\). With the requirement, \(|u_{\mathbf{k}}|^2 - |v_{\mathbf{k}}|^2 = 1\), the transformation diagonalizes the Hamiltonian into \(\mathcal{H} = \sum_{\mathbf{k}} \left( \varepsilon_\alpha \alpha_{\mathbf{k}}^\dagger \alpha_{\mathbf{k}} + \varepsilon_\beta \beta_{\mathbf{k}}^\dagger \beta_{\mathbf{k}} \right) + \text{constant}\). To find the eigenvalues \(\varepsilon_{n}\) and eigenvectors \(u_n = [u_{\mathbf{k}}, v_{\mathbf{k}}]^T\), we require \([ \alpha_{\mathbf{k}}, \mathcal{H} ] = \varepsilon_{\alpha} \alpha_{\mathbf{k}}\), and \([ \beta_{\mathbf{k}}, \mathcal{H} ] = \varepsilon_{\beta} \beta_{\mathbf{k}}\). Using the bosonic commutation relations \([ a_{\mathbf{k}}, a_{\mathbf{k}'}^{\dagger} ] = [ b_{\mathbf{k}}, b_{\mathbf{k}'}^{\dagger} ] = \delta_{\mathbf{k} \mathbf{k}'}\), this leads to the eigenvalue equation, 
\begin{align}
\varepsilon_n \sigma^z u_n(\mathbf{k}) = \mathcal{M}_n(\mathbf{k}) u_n(\mathbf{k})~.\label{eq:eigen_value_equation}
\end{align}
Here, \(\sigma^z\) is the Pauli matrix, and \(\mathcal{M}_n(\mathbf{k})\) is the mode and \(\mathbf{k}\)-dependent matrix. For the \(\alpha\)-mode, we have \(\mathcal{M}_{\alpha}(\mathbf{k}) = \sigma^z \mathcal{H}_{\mathbf{k}}\), and \(\mathcal{M}_{\beta}\) is different.  The matrix \(\mathcal{M}_n\) incorporates the bosonic metric \(\sigma^z\), which accounts for the mixed creation and annihilation operators in \(\psi_{\mathbf{k}}^{\dagger} = [a_{\mathbf{k}}^{\dagger}, ~b_{\mathbf{k}}]\), and it ensures real, positive eigenvalues. 

It is convenient to express $\mathcal{M}_{n}(\mathbf{k})$ in terms of the Pauli spin matrices as, 
\begin{equation} \label{Eq_M}
\mathcal{M}_{n}(\mathbf{k}) = h_0(\mathbf{k}) \sigma^0 + h_x (\mathbf{k})\sigma^x + h_y(\mathbf{k}) \sigma^y + s_n h_z(\mathbf{k}) \sigma^z,
\end{equation}
with  
\( h_0 = J^{\prime} + (\gamma_A + \gamma_B)/2 \), \( h_x = J \text{Re}[\gamma(\mathbf{k})] \), \( h_y = J \text{Im}[\gamma(\mathbf{k})] \) and \( h_z = ( \gamma_A - \gamma_B )/2 \).  For the \(\alpha\)-mode we have \(s_{\alpha} = +1\), and for the \(\beta\)-mode, \(s_{\beta} = -1\) leading to reversal in the sign of \(h_z \sigma^z\) term. Solving for the magnon band energies using \(\det(\mathcal{M}_{n} - \varepsilon_n \sigma^z) = 0\), we obtain,  
\begin{align}
\varepsilon_n = \sqrt{h_0^2 - h_x^2 - h_y^2} + s_n h_z~.
\label{eq:band energies}
\end{align}

From Eq.~(\ref{eq:band energies}), we calculate the spin splitting between two magnon bands to be, 
\begin{align}
\Delta \varepsilon(\mathbf{k}) = \varepsilon_\alpha - \varepsilon_\beta = 2 h_z = \gamma_A(\mathbf{k}) - \gamma_B(\mathbf{k})~.
\end{align}
This yields, 
\begin{align}
\Delta \varepsilon(\mathbf{k}) &= 2S \sum_{m=1}^3 [ (J_{mA} - J_{mB}) \{ \cos(\mathbf{k} \cdot \delta_m) - 1\}  \notag \\
&\quad - 2D^z \sin(\mathbf{k} \cdot \delta_m) ].
\label{eq:spin_splitting}
\end{align}
This dispersive spin splitting, varying across the BZ, arises primarily from the sublattice asymmetry (\( J_{mA} \neq J_{mB} \) and DMI. More importantly, the \(\alpha\)- and \(\beta\)-modes carry opposite spin angular momentum [\( S^z = \sum_{\mathbf{k}} (-\alpha_{\mathbf{k}}^\dagger \alpha_{\mathbf{k}} + \beta_{\mathbf{k}}^\dagger \beta_{\mathbf{k}}) \)]~\cite{Cui_2023}. 

The spin split magnon bands carrying opposite angular momentum generate thermally driven spin currents, positioning insulating altermagnets as a robust platform for magnonic spintronics~\cite{Cheng_2016,Cui_2023}.
\begin{figure*}[t]
    \centering
    \includegraphics[width=0.85\linewidth]{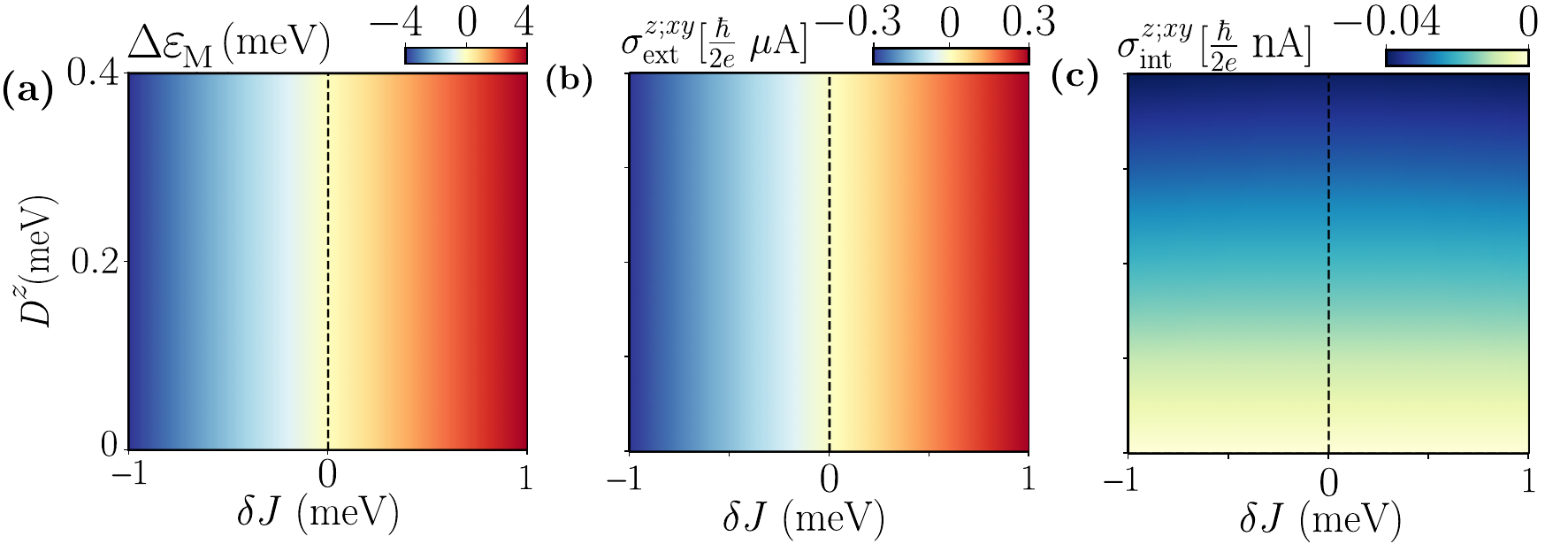}
    \caption{\justifying \textbf{(a)} Phase diagram of the altermagnetic model in the \( (\delta J, D^z) \) parameter space, showing the magnon band splitting at the $M$ point on the BZ boundary. \textbf{(b, c)} Extrinsic and intrinsic spin current conductivities, respectively, plotted over the same parameter space. Parameters are the same as in Fig.~\ref{magnon-bands}, with temperature fixed at \(T = 100\,\mathrm{K}\). 
    \label{delta_J_D_parameter_space}}
\end{figure*}
The color plot of the magnon band dispersion for the $\alpha$-mode in the momentum space is shown in  Fig.~\ref{magnon-bands}\textcolor{blue}{(b)}. In Fig.~\ref{magnon-bands}\textcolor{blue}{(c)}, we present the $\mathbf{k}$-space distribution of the energy splitting between $\alpha$ and $\beta$ modes in absence of DMI. 
But, in presence of both DMI and anisotropic NNN exchange couplings, the spin-splitting of the magnon bands becomes more anisotropic [see Fig.~\ref{magnon-bands}\textcolor{blue}{(f)}]. 

The corresponding eigenvectors, satisfying the orthonormality condition \(\langle u_n | \sigma^z | u_n \rangle = 1\), are identical for both magnon modes and are given by, 
\begin{align}
u_{\alpha} = u_{\beta} = \begin{bmatrix} \cosh \frac{\theta}{2} \\ -\sinh \frac{\theta}{2} e^{i \phi} \end{bmatrix}~.
\label{eigenvectors}
\end{align}
Here, the angles are defined by, \(h_0 = l \cosh \theta\), \(h_x = l \sinh \theta \cos \phi\), \(h_y = l \sinh \theta \sin \phi\). The identical eigenvectors for both magnon bands force them to have identical Berry curvatures. 
For the magnon bands described by Eq.~(\ref{eq:band energies}), the Berry curvature can be expressed as, $\Omega_\alpha^z = \Omega_{\beta}^z$, where 
\begin{align}
\Omega_\alpha^z(\mathbf{k}) = \frac{1}{2} \sinh \theta \left( \frac{\partial \theta}{\partial k_x} \frac{\partial \phi}{\partial k_y} - \frac{\partial \phi}{\partial k_x} \frac{\partial \theta}{\partial k_y} \right)~.
\label{eq:berry_curvature}
\end{align}
The distribution of the Berry curvature in $\mathbf{k}-$space is shown in Fig.~\ref{magnon-bands}\textcolor{blue}{(e)}. We also present band-velocity distribution for $\alpha$-mode in Fig.~\ref{magnon-bands}\textcolor{blue}{(d)}~.
\section{Spin Nernst conductivity} 

Using the framework for magnon-mediated spin transport, developed in Sec \ref{Magnon Spin Current from Thermal Gradient}, we now calculate the spin Nernst conductivity for altermagnetic model, introduced in the previous section. Specifically, we consider a temperature gradient applied along the $\hat{y}$ direction and evaluate the response of the spin-split $\alpha$- and $\beta$-magnon modes to this thermal driving force.

Since these magnon modes carry opposite spin angular momentum and are well-separated in energy [see Eq.~\eqref{eq:spin_splitting}], we treat their contributions to the spin current as independent and additive. Neglecting magnon-magnon interactions, the total spin Nernst response is given by, 
\begin{align}
    \sigma^{z;xy}_{\text{ext}} &= \frac{\hbar ~\tau}{2} \sum_{\mathbf{k}} \left[ \varepsilon_{\beta}~ v^x_{\beta} ~v^y_{\beta} ~\frac{\partial f_{\beta}}{\partial \varepsilon_{\beta}} -  \varepsilon_{\alpha}~ v^x_{\alpha} ~v^y_{\alpha}~ \frac{\partial f_{\alpha}}{\partial \varepsilon_{\alpha}}\right],\\
    \sigma^{z;xy}_{\text{int}} &= \frac{1}{2} \sum_{\mathbf{k}} \left[ \Omega^{z}_{\alpha} (\varepsilon_{\beta} f_{\beta} -  \varepsilon_{\alpha} f_{\alpha}) -2\hbar ( M^z_{\mathbf{\Omega}_{\beta}} - M^z_{\mathbf{\Omega}_{\alpha}}) \right].
\end{align}
Here, $\sigma^{z;xy}_{\text{ext}}$ and $\sigma^{z;xy}_{\text{int}}$ are the extrinsic and intrinsic spin Nernst conductivities, respectively. The extrinsic term originates from scattering-driven population imbalance and scales with $\tau$, while the intrinsic term stems from the Berry curvature and is scattering-independent. The transverse spin current in our model can be used to transfer spin angular momentum to an adjacent ferromagnetic layer, giving rise to SST induced magnetization switching \cite{Karube_2022, Bai_2022, Sayan_2025, Cheng_2018}. 

The Berry curvature-induced orbital magnetization correction $M^z_{\Omega_\nu}$ for band $\nu = (\alpha, \beta)$ is given by~\cite{Matsumoto_2011, Matsumoto_2011_prl}
\begin{equation}
    M^z_{\mathbf{\Omega}_{\nu}} = \frac{1}{\hbar} \int_{\varepsilon_{\nu}(k)}^{\infty} \varepsilon ~\frac{\partial f_{\nu}}{\partial \varepsilon_{\nu}}~ \Omega^z_{\nu}~ d\varepsilon = \frac{k_B T}{\hbar}~g(f_{\nu}) ~\Omega^z_{\nu}~.
\end{equation}
Here, the coefficient $g(f_{\nu})$ is given by,
\begin{equation}
    g(f_{\nu}) = (1 + f_{\nu}) \log(1 + f_{\nu}) - f_{\nu} \log(f_{\nu})~.
\end{equation}
It is a dimensionless weighing function accounting for the thermally occupied states. 
The extrinsic spin currents arise from magnon band splitting due to anisotropic NNN exchange couplings and it vanishes when $\delta J = 0$ in our model. On the other hand, the intrinsic spin currents arise from finite DMI, and they vanish when $D^z = 0$. 
\begin{figure*}[t]
    \centering    \includegraphics[width=\linewidth]{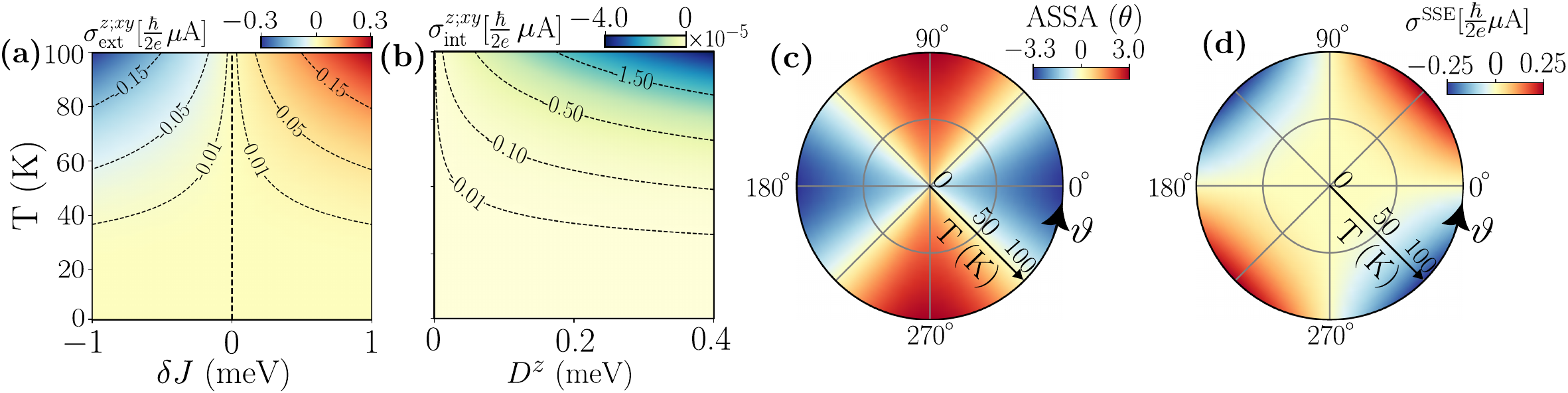}
    \caption{\justifying \textbf{Temperature dependence of the magnon spin conductivities and altermagnetic spin-splitting angle ($\alpha$).} \textbf{(a)} Extrinsic spin Nernst conductivities plotted over the $(\delta J, T)$ parameter space for $\nabla T/T = \partial_y T/T $. \textbf{(b)} Intrinsic spin Nernst conductivities plotted over the $(D^z,T)$ parameter space for same thermal field. The contours show a particular value of the response, scaled by $10^{-5}$ for \textbf{(b)}. \textbf{(c, d)} Polar plots of the altermagnetic spin-splitting angle (ASSA, $\alpha$) and spin Seebeck conductivity, respectively, in the ($\vartheta, ~T$) parameter space. Here, $\vartheta$ denotes the angle between the thermal field direction and the $\hat{x}-$axis. All the parameters for this figure is same as Fig.~\ref{magnon-bands}.}
    \label{ASSA_SSE}
\end{figure*}
While DMI induces spin splitting only near the corners of the Brillouin zone [see Fig.~\ref{magnon-bands}\textcolor{blue}{(g)}], a finite $\delta J$ with $D^z = 0$ results in spin splitting at the midpoint of the symmetry path along the Brillouin zone boundaries, see Fig.~\ref{magnon-bands}\textcolor{blue}{(c)}. Figure~\ref{delta_J_D_parameter_space}\textcolor{blue}{(a)} illustrates the spin splitting $\Delta \varepsilon = \varepsilon_\alpha - \varepsilon_\beta$ at the $M$-point of the Brillouin zone as a function of the parameters $(\delta J, D^z)$. The splitting vanishes for $\delta J = 0$, indicating a transition from the altermagnetic phase to the conventional antiferromagnetic phase with DMI. The corresponding extrinsic spin Nernst conductivity is presented in Fig.~\ref{delta_J_D_parameter_space}\textcolor{blue}{(b)}. It closely follows the behavior of the spin splitting and changes sign with $\delta J$, reflecting the reversal of spin splitting upon changing the sign of $\delta J$. In contrast, the intrinsic spin Nernst conductivity, shown in Fig.~\ref{delta_J_D_parameter_space}\textcolor{blue}{(c)}, is finite only in the presence of DMI.


 
In this study, we assume a low magnon density and apply linear spin wave theory, implying that our analysis is valid at temperatures well below the N\'eel temperature of the antiferromagnetic phase. The temperature dependence of both spin conductivities is shown in Figs.~\ref{ASSA_SSE}\textcolor{blue}{(a)} and \textcolor{blue}{(b)}. Both the extrinsic and intrinsic components increase with temperature, indicating the thermal activation of bosonic magnon modes. 
\section{Spin Seebeck conductivity and Magnon-mediated spin-splitting}

Having demonstrated the magnon spin Nernst effect, we now focus on the associated altermagnetic spin-splitting effect~\cite{Jakub_2021,Karube_2022, Liao, Yuan_2020, Giil_2024} and spin Seebeck effect. While spin-splitting phenomena have been recently predicted and experimentally observed in electronic bands in alternagnetic systems~\cite{Jakub_2021, Karube_2022, Liao}, the corresponding physics in magnon bands is relatively unexplored. In AMs, the non-relativistic ($D^z = 0$) band splitting leads to two magnon modes that propagate at a finite angle to an applied thermal gradient. This gives rise to a transverse spin current, a hallmark of spin-splitting in magnon bands. To characterize the spin-splitting between these two magnon modes, we adopt the \textit{altermagnetic spin-splitting angle} (ASSA or angle $\alpha$), which was originally introduced for electronic bands in altermagnets. For spin current mediated by magnon transport, the ASSA is defined as the angle between the transverse spin current and the longitudinal particle current. It is given by \cite{Jakub_2021}, 
\begin{equation} 
    \text{tan}\frac{\alpha}{2} = \frac{1}{\hbar} \frac{j_s^{z;\perp}}{j_p^{||}} = \frac{(~j^{z;\perp}_{\beta} ~+ j^{z;\perp}_{\alpha}~)}{(~j^{z;||}_{\beta} ~- j^{z;||}_{\alpha}~)} = \frac{(~\sigma^{z;\perp}_{\beta} ~+ \sigma^{z;\perp}_{\alpha}~)}{(~\sigma^{z;||}_{\beta} ~- \sigma^{z;||}_{\alpha}~)}~.\label{ASSA}
\end{equation}
Here, $j^{z;\perp}_s = (j^{z;\perp}_{\beta} + j^{z;\perp}_{\alpha} )$ is the transverse spin current. In Eq.~(\ref{ASSA}), we have  multiplied the particle current by $\hbar$ and expressed it in terms of the spin current to ensure that both quantities have identical dimensions. Since the two magnon modes contribute oppositely to the spin current, the particle current parallel to the thermal field is given by, $j^{||}_p \propto ( j^{z;||}_{\beta} - j^{z;||}_{\alpha} )$. 

To be specific, let us consider a temperature gradient ($\nabla T$) applied along the direction $\hat{n} = [\cos\vartheta\, \hat{x} + \sin\vartheta\, \hat{y}]$. For this configuration, the magnon spin current carried by $\alpha-$mode is given by,
\begin{align} \label{current_1}
    \begin{bmatrix}
        j^{z;||}_{\alpha  } \\ j^{z;\perp}_{\alpha  }
    \end{bmatrix}
    &= \begin{bmatrix}
        \sigma^{z;||}_{\alpha  } \\ \sigma^{z;\perp}_{\alpha  }
    \end{bmatrix}
    \nabla T~ .
\end{align}
Here, $j^{z;||}_{\alpha}$ and $j^{z;\perp}_{\alpha}$ are spin current parallel and perpendicular to the thermal field, respectively and $\sigma^{z;||}_{\alpha}$ and $\sigma^{z;\perp}_{\alpha}$ are corresponding conductivities for the alpha mode. The spin current conductivity $\sigma^{z;||}_{\alpha}$ and $\sigma^{z;\perp}_{\alpha}$ can be expressed in terms of the known response tensors by doing a rotation transformation on  \(\sigma^{z;ab}\) (we derive these in Appendix~\ref{appendix_C}).  The two conductivity components for \(\alpha-\)mode are given by, 
\begin{align}
    \sigma^{z;||}_{\alpha  } &= \sigma^{z;xx}_{\alpha  } \text{cos}^2 \vartheta + \frac{1}{2}  [ \sigma^{z;xy}_{\alpha  } +\sigma^{z;yx}_{\alpha  } ] \text{sin} (2\vartheta)+ \sigma^{z;yy}_{\alpha  } \text{sin}^2 \vartheta, \notag \\
    \sigma^{z;\perp}_{\alpha  } &= \frac{1}{2} [ \sigma^{z;yy}_{\alpha  } -\sigma^{z;xx}_{\alpha  } ] \text{sin} (2 \vartheta) -  \sigma^{z;xy}_{\alpha  } \text{sin}^2 \vartheta + \sigma^{z;yx}_{\alpha  } \text{cos}^2 \vartheta.\label{sigma_perp}
\end{align}
Similar expressions are valid for the $\beta$-mode as well. By substituting Eqs.~\eqref{sigma_perp} into Eq.~\eqref{ASSA}, we obtain a $\vartheta$ dependent ASSA, $\alpha(\vartheta)$. 


In Fig.~\ref{ASSA_SSE}\textcolor{blue}{(c)}, we present a radial color plot showing the dependence of ASSA ($\alpha$) with $\vartheta$ and temperature. It highlights the four-fold symmetry of $\alpha(\vartheta)$ consistent with the energy splitting of the magnon bands in Fig.~\ref{magnon-bands}\textcolor{blue}{(c)}. We observe that $\alpha(\vartheta)$ reaches its maximum when the thermal gradient is applied along the $\hat{x}$ or $\hat{y}$ directions ($\vartheta = m \pi/2$, where $m = 0,1,2,3$). In these cases, spin-up and spin-down magnons make equal and opposite angle with the applied thermal field. This leads to a transverse \textit{pure} spin current.

In contrast, when the thermal field is oriented at intermediate angles, i.e., for $\vartheta \neq  m \pi/2$,  
the spin current of the two magnon modes acquires a finite longitudinal component along with the transverse component. The longitudinal spin current corresponds to the magnon spin Seebeck effect (SSE). The SSE conductivity in the longitudinal direction can be expressed as, 
\begin{equation}
    \sigma^{\text{SSE}} = ( \sigma^{z;||}_{\beta} ~+ \sigma^{z;||}_{\alpha} )~.
\end{equation} 
We show the dependence of SSE conductivity with $\vartheta$ and $T$ in Fig.~\ref{ASSA_SSE}\textcolor{blue}{(d)}. The SSE conductivity also has a four-fold symmetry consistent with energy splitting of the magnon bands. Interestingly, it reaches a maximum at angles where $\alpha(\vartheta)$ is minimum, and vice versa.

\section{Discussion}

We have computed both the energy current and the magnon current in the direction transverse to the applied thermal field for $\nabla T ~|| ~\hat{x} (\hat{y})$. The results indicate that the magnon-mediated spin current is a \textit{pure} spin current, i.e., there is no accompanying energy or magnon particle current in the transverse direction. This establishes our model as a spin-splitter, capable of converting a longitudinal particle current into a \textit{pure} transverse spin current~\cite{Jakub_2021, Yuan_2020, Karube_2022, Giil_2024} for specific directions of $\nabla T$. 

The transverse spin current generated in our model can be effectively utilized to induce SST in an adjacent ferromagnetic layer \cite{Karube_2022, Bai_2022, Sayan_2025, Cheng_2018}. When this pure spin current is injected into the ferromagnet, spin angular momentum is transferred to the local magnetic moments, exerting a spin-transfer torque. This torque can overcome the magnetic anisotropy and trigger magnetization switching. Such all-magnonic spin-split torque mechanisms are central to next-generation spintronic devices, enabling efficient and ultrafast control of magnetic states \cite{Sarma_2004, Bader_2010, chumak_2015}.

To assess the feasibility of an effective SST, we estimate the effective magnetic field generated by the spin current that drives the torque. From Fig.~\ref{delta_J_D_parameter_space}\textcolor{blue}{(b)} we see that, the maximum transverse magnonic spin conductivity is approximately \( \sigma^{z;xy} \approx 0.3~(\hbar/2e)\,\mu\text{A} \), corresponding to a 3D spin conductivity of \( \sigma^{z;xy} \approx 0.03~(\hbar/2e)\,\mu\text{A}/\text{\AA} \). Assuming a temperature gradient of 1~K/nm at \( T = 100~\text{K} \) \cite{Kryder_2008, Cheng_2018}, we get \( \nabla_y T / T = 0.01~\text{\AA}^{-1} \), yielding a maximum spin current density of \( J^{z;x} = 0.64 \times 10^{-5}~\text{J/m}^2 \). When injected into a ferromagnet, this spin current generates a SST per unit magnetization given by \( |\Gamma^s| = \gamma J^{z;x}/(M_s l) \), where \( \gamma \) is the gyromagnetic ratio, \( M_s \) is the saturation magnetization of the ferromagnet, and \( l \) is the ferromagnetic layer thickness \cite{Sinova_2015, Liu_2011}. Using representative values \( l = 5~\text{nm} \) and \( M_s = 6.4 \times 10^5~\text{A/m} \) for a permalloy-based ferromagnet \cite{Torques_2005}, we estimate an effective magnetic field of \( B_{\text{eff}} = |\Gamma^s|/\gamma \approx 2.0~\text{mT} \). This exceeds the typical anisotropy field \(\sim 0.25~\text{mT} \)  for permalloy-based ferromagnets~\cite{Lopusnik_2003, Peng_2019, Husain_2020}, indicating the feasibility of magnon mediated spin-torque-induced switching by altermagnetic insulators.

We now discuss potential material platforms for realizing the model proposed in this study. Recent experimental and \textit{ab initio} studies have identified several hexagonal collinear antiferromagnets that host non-relativistic spin-split electronic bands that exhibit altermagnetic behavior. Notable examples with large spin-splitting and layered structure include \( \text{MnTe, MnSe, MnPSe}_3\), and \(\text{CrSb} \) \cite{Cheng_2019, Lee_2024, Hariki_2024, Sheoran_2024, Reimers_2024}. Additionally, twisted bilayers of \( \text{MnBi}_2\text{Te}_4 \) with antiferromagnetic interlayer interaction also show altermagnetism \cite{Liu_2024_, Pan_2024}. These systems can potentially host spin-split magnon bands and enable spin-splitter torques for magnetization switching. 


\section{Conclusion} 

In this work, we introduced a collinear antiferromagnetic model on a honeycomb lattice with anisotropic NNN exchange couplings and DMI between intra-sublattice sites. This interaction induces altermagnetic spin-split magnon bands without relying on external magnetic fields. In the absence of DMI, the momentum-space spin splitting possesses a four fold symmetry, which becomes more anisotropic on introducing DMI. We investigated magnon-mediated spin transport driven by a thermal gradient and demonstrated the emergence of a \textit{pure} transverse spin current, realizing a magnonic spin-splitter.

Our analysis shows that the Drude-like contribution dominates the spin conductivity, exceeding the Berry-curvature-dependent intrinsic contribution. We further quantified the spin-splitting efficiency using the altermagnetic spin-splitting angle, demonstrating its capability to generate sufficient spin torque in adjacent ferromagnetic layers for magnetization switching. Our work positions magnons in insulating altermagnets as promising candidates for low-dissipation spintronics. 

\section{Acknowledgments}
We thank Sayan Sarkar, Harsh Varshney, and Sunit Das for exciting discussions. S. S. acknowledges Indian Institute of Technology Kanpur for PhD fellowship.  A. A. acknowledges funding from the Core Research Grant by Anusandhan National Research Foundation (ANRF, Sanction No. CRG/2023/007003), Department of Science and Technology, India.  A. A  acknowledges the high-performance computing facility at IIT Kanpur, including HPC 2013, and Param Sanganak.  

\appendix
\section{First order correction to the density matrix}\label{density-matrix-calculations} 
Here, we present the detailed calculation of the equation of motion for the density matrix using the adiabatic switching-on approximation for the temperature gradient. The dynamics of the non-equilibrium density matrix is governed by the quantum Liouville equation~\cite{boyd_2008}, 
\begin{equation}
    \frac{d\rho_{np}}{dt} +\frac{i}{\hbar} [\mathcal{H}, \rho]_{np} = -\frac{i}{\hbar} [\mathcal{H}_{E_T},  \rho]_{np}~.
    \label{A1}
\end{equation}
Here, $\mathcal{H}$ is the unperturbed Bloch Hamiltonian for magnons, and $\mathcal{H}_{E_T}$ is the perturbation due to the thermal field. Expanding the density matrix perturbatively as $\rho = \rho^{(0)} + \rho^{(1)} + \cdots$ in powers of $\nabla T$, we obtain the first-order correction~\cite{boyd_2008}, 
\begin{equation}
    \rho^{\text{(1)}}_{np}(t) = \int^t_{-\infty} -\frac{i}{\hbar} [\mathcal{H}_{E_T}(t'),~ \rho^{(0)}(t')]_{np} ~ e^{i\omega_{np} (t'-t)}~ dt'~.
    \label{A2}
\end{equation}
Here, $\omega_{np} = (\varepsilon_n - \varepsilon_p)/\hbar$. In the main text [see Eq.~(\ref{quantum-kinetic-equation1})], we have identified $-\frac{i}{\hbar} [\mathcal{H}_{E_T}, \rho^{(0)}]_{np} = D_T[\rho^{(0)}]$. Hence, the first-order correction to the density matrix becomes, 
\begin{align}
        \rho^{\text{(1)}}_{np}(t) &= \int^t_{-\infty}-\frac{\tilde{E}^a_T(t')}{2\hbar}\left[\{\mathcal{H}, \partial_a \rho^{(0)}\} - i [\mathcal{R}^a, \{ \mathcal{H}, \rho^{(0)} \}]\right]_{np} \notag \\
        &\quad \times e^{i\omega_{np} (t'-t)} dt'.\label{first_order_density_matrix_B}
\end{align}

To adiabatically switch on the thermal field, we modify the driving frequency as $\omega \rightarrow \omega + i\eta$, where $\eta$ is a small positive parameter. This modifies the thermal driving field to,
$\tilde{E}^a_T(t) = E^a_T ( e^{i\omega t - \eta t} + e^{-i \omega t + \eta t} )/2$. Substituting this into Eq.~(\ref{first_order_density_matrix_B}) we obtain the first order density matrix, 
\begin{align}
    \rho^{(1)}_{np}(t) &= -\frac{1}{4\hbar} E^a_T \left[ \{\mathcal{H}, \partial_a \rho^{(0)}\} - i [\mathcal{R}^a, \{ \mathcal{H}, \rho^{(0)} \}] \right]_{np} \notag \\ &\quad \times \Big\{ \frac{e^{i \omega t}}{\eta + i(\omega_{np} + \omega)} + \frac{e^{-i\omega t}}{\eta + i(\omega_{np} - \omega)} \Big\}\notag \\
    &= -\frac{1}{2\hbar} E^a_T \big[ \varepsilon_n~ \partial_a f_n~ \delta_{np} + i \mathcal{R}^a_{np} ~ \xi_{np} \big] \notag \\ &\quad \times \Big\{ \frac{e^{i \omega t}}{\eta + i(\omega_{np} + \omega)} + \frac{e^{-i\omega t}}{\eta + i(\omega_{np} - \omega)} \Big\}.
    \label{A4}
\end{align}
Here, we have defined $\xi_{np} = [ \varepsilon_n f_n - \varepsilon_pf_p ]$, and $f_n$ is the Bose-Einstein distribution function. Physically, $\eta$ captures the impact of scattering and it is the inverse of the relaxation time, i.e., $\eta = 1/\tau$. 

In the framework of quantum kinetic theory, incorporating relaxation-time effects leads to the following quantum kinetic equation for first order density matrix, 
\begin{equation}
    \frac{\partial \rho^{(1)}_{np}(\mathbf{k},t)}{\partial t} + \frac{i}{\hbar}[\mathcal{H} , \rho^{(1)}(\mathbf{k},t)]_{np} + \frac{\rho^{(1)}_{np}(t)}{\tau} = D_T[\rho^{(0)}]_{np},
    \label{qke_1st_order_correction}
\end{equation}
with the thermal field specified by, $\tilde{E}^a_T(t) = \frac{1}{2} E^a_T (e^{i\omega t} + e^{-i\omega t})$, with $a = (x, y, z)$. This form ensures that relaxation effects are fully incorporated in the linear-response regime. The same result for $\rho^{(1)}_{np}$ can also be obtained by directly solving Eq.~(\ref{qke_1st_order_correction}) with a.c. thermal field driving. 

\section{Magnon Hamiltonian}\label{magnon-bands-derivation}
We derive the magnon Hamiltonian for the altermagnetic spin model introduced in Eq.~\eqref{eq:hamiltonian}. The Hamiltonian captures momentum-dependent spin splitting via anisotropic next-nearest-neighbor (NNN) exchange and Dzyaloshinskii-Moriya interactions (DMI). It can be expressed as, 
\begin{align}
\mathcal{H} &= \mathcal{H}_{\text{NN}} + \mathcal{H}_{\text{NNN}} + \mathcal{H}_{\text{aniso}}~.
\end{align}
Here, \(\mathcal{H}_{\text{NN}}\) is the nearest-neighbor (NN) exchange, \(\mathcal{H}_{\text{NNN}} = \mathcal{H}_{\text{NNN},A} + \mathcal{H}_{\text{NNN},B}\) includes NNN interactions for sublattices \(A\) and \(B\), and \(\mathcal{H}_{\text{aniso}}\) is the uniaxial anisotropy term. The model parameters are presented in the main text. Below, we define each term, apply Holstein-Primakoff (HP) transformations, and transform to momentum space using Fourier transforms, preparing for the magnon dispersion calculation in Section~\ref{sec:spin_model}.

\subsection{Nearest-Neighbor Interaction}

The NN term is given by, 
\begin{align}
\mathcal{H}_{\text{NN}} &= J \sum_{\langle i \in A, j \in B \rangle} \mathbf{S}_i \cdot \mathbf{S}_j \notag \\
&= J \sum_{\langle i \in A, j \in B \rangle} \left[ S_i^z S_j^z + \frac{1}{2} (S_i^+ S_j^- + S_i^- S_j^+) \right]~.
\label{eq:nn-spin}
\end{align}
Here, \(S_i^\pm = S_i^x \pm i S_i^y\) are spin raising/lowering operators. We apply HP transformations for a collinear antiferromagnet with spins aligned along \(\pm z\). Using the HP transformations, the NN Hamiltonian becomes, 
\begin{equation}
    \mathcal{H}_{NN} = JS \sum_{\langle i \in A, j \in B \rangle} [ a_i^{\dagger} a_i + b_j^{\dagger} b_j + a_i b_j + a_i^{\dagger} b_j^{\dagger}]~.
    \label{ab_nearst-neighbor-hamiltonian}
\end{equation}
In the above simplifications, we have dropped the constant term and ignored magnon-magnon interactions. To convert real space operators to momentum space operators, we use the Fourier transformation [see Eq.~(\ref{Fourier_transformation})]. Applying the Fourier transformations and using the identity $\delta(\mathbf{k} - \mathbf{k}') = 1/N \sum_i \text{exp}[i (\mathbf{k} - \mathbf{k}') \cdot \mathbf{r}_i]$, we obtain, 
\begin{align}
    \mathcal{H}_{NN} &= JS \sum_{\mathbf{k}} [ 3( a^{\dagger}_{\mathbf{k}} a_{\mathbf{k}} + b^{\dagger}_{\mathbf{k}} b_{\mathbf{k}}) + \gamma(\mathbf{k})^* a_{\mathbf{k}} b_{\mathbf{k}} + \gamma(\mathbf{k}) a^{\dagger}_{\mathbf{k}} b^{\dagger}_{\mathbf{k}} ].
\end{align}
Here, $\gamma(\mathbf{k})$ denotes the structure factor for NN exchange interaction, and it is defined as $ \gamma(\mathbf{k}) = \sum_{\mathbf{d}_j} \text{exp}[ i\mathbf{k} \cdot \mathbf{d}_j] $. From Fig.~\ref{magnon-bands}\textcolor{blue}{(a)}, we identify the NN position vectors to be,  
\begin{equation*}
    \mathbf{d}_1 = (\frac{\sqrt{3}a}{2}, \frac{a}{2}),~ \mathbf{d}_2 =  ( -\frac{\sqrt{3}a}{2}, \frac{a}{2} ),~ \text{and}~~ \mathbf{d}_3 = ( 0, -a )~.
\end{equation*}
Using these, the structure factor becomes, 
\begin{align}
    \gamma(\mathbf{k}) = 2 e^{iak_y/2} ~\text{cos}( \frac{\sqrt{3} a k_x}{2} ) + e^{-i a k_y}~.\label{eq:gamma-k}
\end{align}

\subsection{Next-Nearest-Neighbor Interaction and DMI} 
The NNN term can be expressed as 
\begin{align}
\mathcal{H}_{\text{NNN}} = \mathcal{H}_{\text{NNN},A} + \mathcal{H}_{\text{NNN},B}~, 
\end{align}
with 
\begin{align}
\mathcal{H}_{\text{NNN},A} &= \frac{1}{2} \sum_{m=1}^6 \sum_{i \in A} \left[ J_{mA} \mathbf{S}_i \cdot \mathbf{S}_{i + \delta_m} + \mathbf{D}_{mA} \cdot (\mathbf{S}_i \times \mathbf{S}_{i + \delta_m}) \right], \notag \\
\mathcal{H}_{\text{NNN},B} &= 
\frac{1}{2} \sum_{m=1}^6 \sum_{i \in B} \left[ J_{mB} \mathbf{S}_i \cdot \mathbf{S}_{i + \delta_m'} + \mathbf{D}_{mB} \cdot (\mathbf{S}_i \times \mathbf{S}_{i + \delta_m'}) \right]. 
\end{align}

In the main text, we have assumed that the NNN exchange couplings satisfies the following relations: for sublattice A,  
$J_{1A} = J_{4A} = J_1$, $J_{2A} = J_{5A} = J_1 - \delta J$, and $J_{3A} = J_{6A} = J_1 + \delta J$ and for sublattice B, $J_{1B} = J_{4B} = J_1 - \delta J$, $J_{2B} = J_{5B} = J_1$, and $J_{3B} = J_{6B} = J_1 + \delta J$. On the other hand the DMI vector satisfies the following conditions: $\mathbf{D}_{mA} = +D^z \hat{z}$ for NNN position vectors $\delta_{1,2,3}$ and $\mathbf{D}_{mA} = -D^z \hat{z}$ for NNN position vectors $\delta_{4,5,6}$. With that assumptions and the fact that $\delta_{4,5,6} =- \delta_{1,2,3}$, in the following calculations for deriving the magnon Hamiltonian, we consider only three NNN connected by $\delta_{1,2,3}$ and omit the factor $\frac{1}{2}$. For sublattice \(A\), with \(\mathbf{D}_{mA} = +D^z \hat{z}\) for clockwise bonds (\(\delta_{1,2,3}\)), we have,
\begin{align}
\mathbf{S}_i \cdot \mathbf{S}_{i + \delta_m} &= S_i^z S_{i + \delta_m}^z + \frac{1}{2} (S_i^+ S_{i + \delta_m}^- + S_i^- S_{i + \delta_m}^+), \notag \\
\mathbf{D}_{mA} \cdot (\mathbf{S}_i \times \mathbf{S}_{i + \delta_m}) &= 
- \frac{i D^z}{2} (S_i^+ S_{i + \delta_m}^- - S_i^- S_{i + \delta_m}^+).
\end{align}
Using this we obtain, 
\begin{widetext}
\begin{align}
\mathcal{H}_{\text{NNN},A} = \sum_{m=1}^3 \sum_{i \in A} \left[ J_{mA} \left( S_i^z S_{i + \delta_m}^z + \frac{1}{2} (S_i^+ S_{i + \delta_m}^- + S_i^- S_{i + \delta_m}^+) \right) - \frac{i D^z}{2} (S_i^+ S_{i + \delta_m}^- - S_i^- S_{i + \delta_m}^+) \right].
\label{eq:nnn-a}
\end{align}

Applying HP transformations, the exchange term becomes,
\begin{align}
J_{mA} \mathbf{S}_i \cdot \mathbf{S}_{i + \delta_m} \approx J_{mA} \left[ -S^2 - S (a_i^\dagger a_i + a_{i + \delta_m}^\dagger a_{i + \delta_m}) + S (a_i a_{i + \delta_m}^\dagger + a_i^\dagger a_{i + \delta_m}) \right].
\end{align}
Similarly, the DMI term reduces to, 
\begin{align}
- \frac{i D^z}{2} (S_i^+ S_{i + \delta_m}^- - S_i^- S_{i + \delta_m}^+) \approx - i D^z S (a_i a_{i + \delta_m}^\dagger - a_i^\dagger a_{i + \delta_m})~.
\end{align}
Combining these equations and dropping constants, we obtain 
\begin{align}
\mathcal{H}_{\text{NNN},A} = S \sum_{m=1}^3 \sum_{i \in A} \left[ J_{mA} (a_i a_{i + \delta_m}^\dagger + a_i^\dagger a_{i + \delta_m}) - J_{mA} (a_i^\dagger a_i + a_{i + \delta_m}^\dagger a_{i + \delta_m}) - i D^z (a_i a_{i + \delta_m}^\dagger - a_i^\dagger a_{i + \delta_m}) \right]~.
\end{align}
Using the Fourier transform, this reduces to  
\begin{align}
\sum_{i \in A} a_i a_{i + \delta_m}^\dagger = \frac{1}{N} \sum_{i \in A} \sum_{\mathbf{k}, \mathbf{k}'} e^{i \mathbf{k} \cdot \mathbf{r}_i} e^{-i \mathbf{k}' \cdot (\mathbf{r}_i + \delta_m)} a_{\mathbf{k}} a_{\mathbf{k}'}^\dagger = \sum_{\mathbf{k}} e^{-i \mathbf{k} \cdot \delta_m} a_{\mathbf{k}} a_{\mathbf{k}}^\dagger~.
\end{align}

Summing over \(m=1,3\), results in  
\begin{align}
\mathcal{H}_{\text{NNN},A} = 2S \sum_{\mathbf{k}} \sum_{m=1}^3 \left[ J_{mA} \left( \cos(\mathbf{k} \cdot \delta_m) - 1 \right) - D^z \sin(\mathbf{k} \cdot \delta_m) \right] a_{\mathbf{k}}^\dagger a_{\mathbf{k}}~.
\label{eq:nnn-a-k}
\end{align}
\end{widetext}
For a similar calculation on \(\mathcal{H}_{\text{NNN},B}\), we substitute \(A \to B\) and replace \(J_{mA} \to J_{mB}\), \(\delta_m \to \delta_m' = -\delta_m\), and \(a_{\mathbf{k}} \to b_{\mathbf{k}}\). Since \(\cos(-\mathbf{k} \cdot \delta_m) = \cos(\mathbf{k} \cdot \delta_m)\), \(\sin(-\mathbf{k} \cdot \delta_m) = -\sin(\mathbf{k} \cdot \delta_m)\), we have 
\begin{align}
\mathcal{H}_{\text{NNN},B} &= 2S \sum_{\mathbf{k}} \sum_{m=1}^3 [ J_{mB} \left( \cos(\mathbf{k} \cdot \delta_m) - 1 \right) \notag \\
&\quad + D^z \sin(\mathbf{k} \cdot \delta_m) ] b_{\mathbf{k}}^\dagger b_{\mathbf{k}}~.
\label{eq:nnn-b-k}
\end{align}

\subsection{Anisotropy Term}

The anisotropy term simplifies to, 
\begin{align}
    \mathcal{H}_{\text{aniso}} &= K \sum_{i} [ (S^z_{iA})^2 + (S^z_{iB})^2 ] \notag \\
    &= -2SK \sum_{\mathbf{k}} [ a_{\mathbf{k}}^{\dagger} a_{\mathbf{k}} + b_{\mathbf{k}}^{\dagger} b_{\mathbf{k}} ]~.
\end{align}
Here, $K<0$ ensures a collinear antiferromagnetic ground state. 

\subsection{Total Momentum-Space Hamiltonian}

Combining all the terms, $\mathcal{H} = \mathcal{H}_{NN} + \mathcal{H}_{NNN} + \mathcal{H}_{\rm aniso} = \sum_{\mathbf{k}} \psi_{\mathbf{k}}^{\dagger} ~\mathcal{H}_{\mathbf{k}} ~\psi_{\mathbf{k}}$, where $\psi_{\mathbf{k}}^{\dagger} = [ a_{\mathbf{k}}^{\dagger}~~~~b_{\mathbf{k}}]$. The momentum resolved matrix Hamiltonian $\mathcal{H}_{\mathbf{k}}$ is given by, 
\begin{align}
\mathcal{H}_{\mathbf{k}} = \begin{pmatrix}
J' + \gamma_A(\mathbf{k}) & JS \gamma(\mathbf{k}) \\
JS \gamma(\mathbf{k})^* & J' + \gamma_B(\mathbf{k})
\end{pmatrix},
\end{align}
with \(J' = 3JS - 2KS\). The structure factors are, 
\begin{align}
\gamma_A(\mathbf{k}) &= 2S \sum_{m=1}^3 \left[ J_{mA} \left( \cos(\mathbf{k} \cdot \delta_m) - 1 \right) - D^z \sin(\mathbf{k} \cdot \delta_m) \right], \notag \\
\gamma_B(\mathbf{k}) &= 2S \sum_{m=1}^3 \left[ J_{mB} \left( \cos(\mathbf{k} \cdot \delta_m) - 1 \right) + D^z \sin(\mathbf{k} \cdot \delta_m) \right],
\end{align}
and \(\gamma(\mathbf{k})\) from Eq.~\eqref{eq:gamma-k}. Note that this is a non-Hermitian matrix due to the mixed operator basis. Therefore, it requires a Bogoliubov transformation to obtain the magnon dispersion, as discussed in Section~\ref{sec:spin_model}.

\subsection{Similarity to the Haldane Model for electrons}
\label{app:haldane-model}
In this subsection, we reinterpret the spin Hamiltonian in Eq.~\eqref{eq:hamiltonian} as a spin analog of the Haldane model for electrons~\cite{Haldane_1998}. In the spin model, the anisotropic NNN exchange and DMI introduce complex phase factors analogous to complex direction-dependent hopping terms in the Haldane model. 

To understand this, let us focus on the NNN terms for sublattice \(A\). In terms of the spin raising and lowering operators, it is given by Eq.~\eqref{eq:nnn-a} which can be rewritten as 
\begin{widetext}
\begin{align}
\mathcal{H}_{\text{NNN},A} = \frac{1}{2} \sum_{m=1}^6 \sum_{i \in A} \left[ J_{mA} S_i^z S_{i + \delta_m}^z + \frac{1}{2} \left( J_{mA} - i D^z \right) S_i^+ S_{i + \delta_m}^- + \frac{1}{2} \left( J_{mA} + i D^z \right) S_i^- S_{i + \delta_m}^+ \right].
\label{eq:nnn-a-spin}
\end{align}
We define the effective NNN coupling in polar form, 
\begin{align}
\tilde{J}_{mA} e^{i \phi_{mA}} = J_{mA} - i D^z, \quad \tilde{J}_{mA} e^{-i \phi_{mA}} = J_{mA} + i D^z,
\end{align}
Note that \(\phi_{mA}\) depends on the sign of \(D^z\), which is positive for clockwise bonds as shown in Fig.~\ref{magnon-bands}\textcolor{blue}{(a)}. Thus, we have
\begin{align}
\mathcal{H}_{\text{NNN},A} = \sum_{m=1}^6 \sum_{i \in A} \left[ J_{mA} S_i^z S_{i + \delta_m}^z + \frac{1}{2} \tilde{J}_{mA} \left( e^{i \phi_{mA}} S_i^+ S_{i + \delta_m}^- + e^{-i \phi_{mA}} S_i^- S_{i + \delta_m}^+ \right) \right]~.
\label{eq:nnn-a-haldane}
\end{align}

A similar calculation for sublattice \(B\) yields 
\begin{align}
\mathcal{H}_{\text{NNN},B} = \sum_{m=1}^6 \sum_{i \in B} \left[ J_{mB} S_i^z S_{i + \delta_m'}^z + \frac{1}{2} \tilde{J}_{mB} \left( e^{i \phi_{mB}} S_i^+ S_{i + \delta_m'}^- + e^{-i \phi_{mB}} S_i^- S_{i + \delta_m'}^+ \right) \right],
\end{align}
with \(\tilde{J}_{mB} = \sqrt{J_{mB}^2 + (D^z)^2}\), \(\phi_{mB} = \tan^{-1}\left(- \frac{D^z}{J_{mB}} \right)\).
\end{widetext}
This form resembles the Haldane model, where the complex phases \(\phi_{mA}\), \(\phi_{mB}\) in the NNN terms mimic the imaginary hopping terms that give rise to finite Berry curvature and chern phases in the original Haldane model~\cite{Haldane_1998}. In the context of altermagnets, the anisotropic \(J_{mA}\), \(J_{mB}\) are sufficient to break the magnon band degeneracy and give rise to momentum-dependent spin splitting. However, the DMI interactions are crucial to break the effective time reversal symmetry, which makes the intrinsic spin response finite for this model. 

\section{Spin current components parallel and transverse to the thermal field}\label{appendix_C}
\begin{figure}[h]
    \centering
    \includegraphics[width=0.4\linewidth]{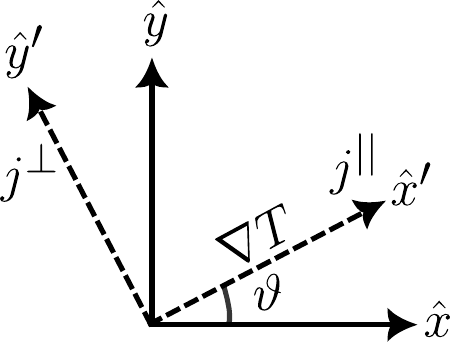}
    \caption{\justifying Schematic of the original coordinate system ($x, y$) and the rotated frame ($x', y'$) obtained by a counterclockwise rotation by angle $\vartheta$, such that $\hat{x}^{\prime} || \nabla T$. Here, $j^{||}$ and $j^{\perp}$ are spin currents along and transverse to the thermal field direction, respectively.}
    \label{lab_frame_rotated_frame}
\end{figure}
In this section, we derive the spin current components parallel (\( j^{||} \)) and perpendicular (\( j^{\perp} \)) to the applied thermal field. The spin current response in the lab frame (without explicit magnon band indices) can be expressed as, 
\begin{align}
    \begin{bmatrix}
        j^{z;x} \\ j^{z;y}
    \end{bmatrix}
     = 
     \begin{bmatrix}
         \sigma^{z;xx} & \sigma^{z;xy} \\
         \sigma^{z;yx} & \sigma^{z;yy}
     \end{bmatrix}
     \begin{bmatrix}
         \nabla_x T \\ \nabla_y T
     \end{bmatrix} \Rightarrow \mathbf{j} = \sigma \cdot \nabla T~.\label{field-response-relation}
\end{align}
Following Fig.~\ref{lab_frame_rotated_frame}, we introduce a rotated frame in which the thermal field points along the \( x^{\prime} \)-axis. The spin current and gradient vectors transform under rotation as, 
\begin{equation}
    \mathbf{j}^{\prime} = R_z(\vartheta) \cdot \mathbf{j}, \quad \nabla^{\prime} T = R_z(\vartheta) \cdot \nabla T~. \label{sigma-rotated-frame}
\end{equation}
Here, \( R_z(\vartheta) \) is the counterclockwise rotation matrix for angle $\vartheta$. It is given by, 
\begin{equation}
R_z(\vartheta) = 
\begin{bmatrix}
\cos\vartheta & \sin\vartheta \\
-\sin\vartheta & \cos\vartheta
\end{bmatrix}~.
\end{equation}

Since the thermal field is aligned along the \( x^{\prime} \)-direction, we have \( \nabla^{\prime} T = [\nabla_{x^{\prime}}T, 0]^T \), and the spin current components in the rotated from are given by, 
\begin{align}
\mathbf{j}^{\prime} = \bm{\sigma}^{\prime} \cdot \nabla^{\prime} T~.
\end{align}
The conductivity tensor in the rotated frame is obtained by the standard transformation, 
\begin{align}
    \bm{\sigma}^{\prime} = R_z(\vartheta) \cdot \bm{\sigma} \cdot R_z^{T}(\vartheta)~.
    \label{conductivity-rotation}
\end{align}
Combining these equations, the parallel and transverse spin current components reduce to, 
\begin{align}
    \begin{bmatrix}
        j^{||} \\ j^{\perp}
    \end{bmatrix}
     = 
     \begin{bmatrix}
         j^{z;x^{\prime}} \\ j^{z;y^{\prime}}
     \end{bmatrix}
     = 
     \begin{bmatrix}
         \sigma^{z;x^{\prime}x^{\prime}} \\ \sigma^{z;y^{\prime}x^{\prime}} 
     \end{bmatrix} \nabla_{x^{\prime}}T~.
    \label{j_parallel_j_perp}
\end{align}
Using Eq.~(\ref{conductivity-rotation}), the longitudinal and transverse spin conductivities with respect to the thermal field are given by, 
\begin{align}
     \sigma^{||} & = \sigma^{z;x^{\prime} x^{\prime}} \notag \\
    &= \sigma^{z;xx} \cos^2\vartheta +  \frac{1}{2}(\sigma^{z;xy}+\sigma^{z;yx}) \sin (2\vartheta) + \sigma^{z;yy} \sin^2\vartheta, \notag\\
    \sigma^{\perp} &= \sigma^{z;y^{\prime} x^{\prime}} \notag\\
    &= \frac{1}{2}(\sigma^{z;yy} - \sigma^{z;xx}) \sin(2\vartheta) -  \sigma^{z;xy} \sin^2\vartheta + \sigma^{z;yx} \cos^2\vartheta.
    \label{sigma_parallel_parpendicular}
\end{align}

\bibliography{references}

\begin{thebibliography}{71}%
\makeatletter
\providecommand \@ifxundefined [1]{%
 \@ifx{#1\undefined}
}%
\providecommand \@ifnum [1]{%
 \ifnum #1\expandafter \@firstoftwo
 \else \expandafter \@secondoftwo
 \fi
}%
\providecommand \@ifx [1]{%
 \ifx #1\expandafter \@firstoftwo
 \else \expandafter \@secondoftwo
 \fi
}%
\providecommand \natexlab [1]{#1}%
\providecommand \enquote  [1]{``#1''}%
\providecommand \bibnamefont  [1]{#1}%
\providecommand \bibfnamefont [1]{#1}%
\providecommand \citenamefont [1]{#1}%
\providecommand \href@noop [0]{\@secondoftwo}%
\providecommand \href [0]{\begingroup \@sanitize@url \@href}%
\providecommand \@href[1]{\@@startlink{#1}\@@href}%
\providecommand \@@href[1]{\endgroup#1\@@endlink}%
\providecommand \@sanitize@url [0]{\catcode `\\12\catcode `\$12\catcode
  `\&12\catcode `\#12\catcode `\^12\catcode `\_12\catcode `\%12\relax}%
\providecommand \@@startlink[1]{}%
\providecommand \@@endlink[0]{}%
\providecommand \url  [0]{\begingroup\@sanitize@url \@url }%
\providecommand \@url [1]{\endgroup\@href {#1}{\urlprefix }}%
\providecommand \urlprefix  [0]{URL }%
\providecommand \Eprint [0]{\href }%
\providecommand \doibase [0]{http://dx.doi.org/}%
\providecommand \selectlanguage [0]{\@gobble}%
\providecommand \bibinfo  [0]{\@secondoftwo}%
\providecommand \bibfield  [0]{\@secondoftwo}%
\providecommand \translation [1]{[#1]}%
\providecommand \BibitemOpen [0]{}%
\providecommand \bibitemStop [0]{}%
\providecommand \bibitemNoStop [0]{.\EOS\space}%
\providecommand \EOS [0]{\spacefactor3000\relax}%
\providecommand \BibitemShut  [1]{\csname bibitem#1\endcsname}%
\let\auto@bib@innerbib\@empty
\bibitem [{\citenamefont {\ifmmode \check{Z}\else
  \v{Z}\fi{}uti\ifmmode~\acute{c}\else \'{c}\fi{}}\ \emph
  {et~al.}(2004)\citenamefont {\ifmmode \check{Z}\else
  \v{Z}\fi{}uti\ifmmode~\acute{c}\else \'{c}\fi{}}, \citenamefont {Fabian},\
  and\ \citenamefont {Das~Sarma}}]{Sarma_2004}%
  \BibitemOpen
  \bibfield  {author} {\bibinfo {author} {\bibfnamefont {Igor}\ \bibnamefont
  {\ifmmode \check{Z}\else \v{Z}\fi{}uti\ifmmode~\acute{c}\else \'{c}\fi{}}},
  \bibinfo {author} {\bibfnamefont {Jaroslav}\ \bibnamefont {Fabian}}, \ and\
  \bibinfo {author} {\bibfnamefont {S.}~\bibnamefont {Das~Sarma}},\ }\bibfield
  {title} {\enquote {\bibinfo {title} {Spintronics: Fundamentals and
  applications},}\ }\href {\doibase 10.1103/RevModPhys.76.323} {\bibfield
  {journal} {\bibinfo  {journal} {Rev. Mod. Phys.}\ }\textbf {\bibinfo {volume}
  {76}},\ \bibinfo {pages} {323--410} (\bibinfo {year} {2004})}\BibitemShut
  {NoStop}%
\bibitem [{\citenamefont {Bader}\ and\ \citenamefont
  {Parkin}(2010)}]{Bader_2010}%
  \BibitemOpen
  \bibfield  {author} {\bibinfo {author} {\bibfnamefont {S.D.}\ \bibnamefont
  {Bader}}\ and\ \bibinfo {author} {\bibfnamefont {S.S.P.}\ \bibnamefont
  {Parkin}},\ }\bibfield  {title} {\enquote {\bibinfo {title} {Spintronics},}\
  }\href {\doibase https://doi.org/10.1146/annurev-conmatphys-070909-104123}
  {\bibfield  {journal} {\bibinfo  {journal} {Annual Review of Condensed Matter
  Physics}\ }\textbf {\bibinfo {volume} {1}},\ \bibinfo {pages} {71--88}
  (\bibinfo {year} {2010})}\BibitemShut {NoStop}%
\bibitem [{\citenamefont {\textit{et al.}}(2020)}]{hirohata_2020}%
  \BibitemOpen
  \bibfield  {author} {\bibinfo {author} {\bibfnamefont {Atsufumi~Hirohata}\
  \bibnamefont {\textit{et al.}}},\ }\bibfield  {title} {\enquote {\bibinfo
  {title} {Review on spintronics: Principles and device applications},}\ }\href
  {https://www.sciencedirect.com/science/article/pii/S0304885320302353}
  {\bibfield  {journal} {\bibinfo  {journal} {Journal of Magnetism and Magnetic
  Materials}\ }\textbf {\bibinfo {volume} {509}},\ \bibinfo {pages} {166711}
  (\bibinfo {year} {2020})}\BibitemShut {NoStop}%
\bibitem [{\citenamefont {Maekawa}\ \emph {et~al.}(2012)\citenamefont
  {Maekawa}, \citenamefont {Valenzuela}, \citenamefont {Saitoh},\ and\
  \citenamefont {Kimura}}]{Maekawa_2012}%
  \BibitemOpen
  \bibfield  {author} {\bibinfo {author} {\bibfnamefont {Sadamichi}\
  \bibnamefont {Maekawa}}, \bibinfo {author} {\bibfnamefont {Sergio~O.}\
  \bibnamefont {Valenzuela}}, \bibinfo {author} {\bibfnamefont {Eiji}\
  \bibnamefont {Saitoh}}, \ and\ \bibinfo {author} {\bibfnamefont {Takashi}\
  \bibnamefont {Kimura}},\ }\href
  {https://doi.org/10.1093/acprof:oso/9780199600380.001.0001} {\emph {\bibinfo
  {title} {Spin Current}}}\ (\bibinfo  {publisher} {Oxford University Press},\
  \bibinfo {year} {2012})\BibitemShut {NoStop}%
\bibitem [{\citenamefont {Blundell}(2001)}]{blundell_2001}%
  \BibitemOpen
  \bibfield  {author} {\bibinfo {author} {\bibfnamefont {S.}~\bibnamefont
  {Blundell}},\ }\href {https://books.google.co.in/books?id=zP9QEAAAQBAJ}
  {\emph {\bibinfo {title} {Magnetism in Condensed Matter}}},\ Oxford Master
  Series in Condensed Matter Physics\ (\bibinfo  {publisher} {OUP Oxford},\
  \bibinfo {year} {2001})\BibitemShut {NoStop}%
\bibitem [{\citenamefont {Baltz}\ \emph {et~al.}(2018)\citenamefont {Baltz},
  \citenamefont {Manchon}, \citenamefont {Tsoi}, \citenamefont {Moriyama},
  \citenamefont {Ono},\ and\ \citenamefont {Tserkovnyak}}]{Baltz_2018}%
  \BibitemOpen
  \bibfield  {author} {\bibinfo {author} {\bibfnamefont {V.}~\bibnamefont
  {Baltz}}, \bibinfo {author} {\bibfnamefont {A.}~\bibnamefont {Manchon}},
  \bibinfo {author} {\bibfnamefont {M.}~\bibnamefont {Tsoi}}, \bibinfo {author}
  {\bibfnamefont {T.}~\bibnamefont {Moriyama}}, \bibinfo {author}
  {\bibfnamefont {T.}~\bibnamefont {Ono}}, \ and\ \bibinfo {author}
  {\bibfnamefont {Y.}~\bibnamefont {Tserkovnyak}},\ }\bibfield  {title}
  {\enquote {\bibinfo {title} {Antiferromagnetic spintronics},}\ }\href
  {\doibase 10.1103/RevModPhys.90.015005} {\bibfield  {journal} {\bibinfo
  {journal} {Rev. Mod. Phys.}\ }\textbf {\bibinfo {volume} {90}},\ \bibinfo
  {pages} {015005} (\bibinfo {year} {2018})}\BibitemShut {NoStop}%
\bibitem [{\citenamefont {Chumak}\ \emph {et~al.}(2015)\citenamefont {Chumak},
  \citenamefont {Vasyuchka}, \citenamefont {Serga},\ and\ \citenamefont
  {Hillebrands}}]{chumak_2015}%
  \BibitemOpen
  \bibfield  {author} {\bibinfo {author} {\bibfnamefont {Andrii~V}\
  \bibnamefont {Chumak}}, \bibinfo {author} {\bibfnamefont {Vitaliy~I}\
  \bibnamefont {Vasyuchka}}, \bibinfo {author} {\bibfnamefont {Alexander~A}\
  \bibnamefont {Serga}}, \ and\ \bibinfo {author} {\bibfnamefont {Burkard}\
  \bibnamefont {Hillebrands}},\ }\bibfield  {title} {\enquote {\bibinfo {title}
  {Magnon spintronics},}\ }\href
  {https://www.nature.com/articles/nphys3347#citeas} {\bibfield  {journal}
  {\bibinfo  {journal} {Nature physics}\ }\textbf {\bibinfo {volume} {11}},\
  \bibinfo {pages} {453--461} (\bibinfo {year} {2015})}\BibitemShut {NoStop}%
\bibitem [{\citenamefont {Cheng}\ \emph {et~al.}(2016)\citenamefont {Cheng},
  \citenamefont {Okamoto},\ and\ \citenamefont {Xiao}}]{Cheng_2016}%
  \BibitemOpen
  \bibfield  {author} {\bibinfo {author} {\bibfnamefont {Ran}\ \bibnamefont
  {Cheng}}, \bibinfo {author} {\bibfnamefont {Satoshi}\ \bibnamefont
  {Okamoto}}, \ and\ \bibinfo {author} {\bibfnamefont {Di}~\bibnamefont
  {Xiao}},\ }\bibfield  {title} {\enquote {\bibinfo {title} {Spin nernst effect
  of magnons in collinear antiferromagnets},}\ }\href {\doibase
  10.1103/PhysRevLett.117.217202} {\bibfield  {journal} {\bibinfo  {journal}
  {Phys. Rev. Lett.}\ }\textbf {\bibinfo {volume} {117}},\ \bibinfo {pages}
  {217202} (\bibinfo {year} {2016})}\BibitemShut {NoStop}%
\bibitem [{\citenamefont {Zyuzin}\ and\ \citenamefont
  {Kovalev}(2016)}]{Vladimir_2016}%
  \BibitemOpen
  \bibfield  {author} {\bibinfo {author} {\bibfnamefont {Vladimir~A.}\
  \bibnamefont {Zyuzin}}\ and\ \bibinfo {author} {\bibfnamefont {Alexey~A.}\
  \bibnamefont {Kovalev}},\ }\bibfield  {title} {\enquote {\bibinfo {title}
  {Magnon spin nernst effect in antiferromagnets},}\ }\href {\doibase
  10.1103/PhysRevLett.117.217203} {\bibfield  {journal} {\bibinfo  {journal}
  {Phys. Rev. Lett.}\ }\textbf {\bibinfo {volume} {117}},\ \bibinfo {pages}
  {217203} (\bibinfo {year} {2016})}\BibitemShut {NoStop}%
\bibitem [{\citenamefont {Rezende}\ \emph {et~al.}(2019)\citenamefont
  {Rezende}, \citenamefont {Azevedo},\ and\ \citenamefont
  {Rodríguez-Suárez}}]{Rezende_2019}%
  \BibitemOpen
  \bibfield  {author} {\bibinfo {author} {\bibfnamefont {Sergio~M.}\
  \bibnamefont {Rezende}}, \bibinfo {author} {\bibfnamefont {Antonio}\
  \bibnamefont {Azevedo}}, \ and\ \bibinfo {author} {\bibfnamefont
  {Roberto~L.}\ \bibnamefont {Rodríguez-Suárez}},\ }\bibfield  {title}
  {\enquote {\bibinfo {title} {Introduction to antiferromagnetic magnons},}\
  }\href {\doibase 10.1063/1.5109132} {\bibfield  {journal} {\bibinfo
  {journal} {Journal of Applied Physics}\ }\textbf {\bibinfo {volume} {126}},\
  \bibinfo {pages} {151101} (\bibinfo {year} {2019})}\BibitemShut {NoStop}%
\bibitem [{\citenamefont {Rezende}\ \emph {et~al.}(2016)\citenamefont
  {Rezende}, \citenamefont {Rodr\'{\i}guez-Su\'arez},\ and\ \citenamefont
  {Azevedo}}]{Rezende_2016}%
  \BibitemOpen
  \bibfield  {author} {\bibinfo {author} {\bibfnamefont {S.~M.}\ \bibnamefont
  {Rezende}}, \bibinfo {author} {\bibfnamefont {R.~L.}\ \bibnamefont
  {Rodr\'{\i}guez-Su\'arez}}, \ and\ \bibinfo {author} {\bibfnamefont
  {A.}~\bibnamefont {Azevedo}},\ }\bibfield  {title} {\enquote {\bibinfo
  {title} {Theory of the spin seebeck effect in antiferromagnets},}\ }\href
  {\doibase 10.1103/PhysRevB.93.014425} {\bibfield  {journal} {\bibinfo
  {journal} {Phys. Rev. B}\ }\textbf {\bibinfo {volume} {93}},\ \bibinfo
  {pages} {014425} (\bibinfo {year} {2016})}\BibitemShut {NoStop}%
\bibitem [{\citenamefont {Shen}(2019)}]{Shen_2019}%
  \BibitemOpen
  \bibfield  {author} {\bibinfo {author} {\bibfnamefont {Ka}~\bibnamefont
  {Shen}},\ }\bibfield  {title} {\enquote {\bibinfo {title} {Pure spin current
  in antiferromagnetic insulators},}\ }\href {\doibase
  10.1103/PhysRevB.100.094423} {\bibfield  {journal} {\bibinfo  {journal}
  {Phys. Rev. B}\ }\textbf {\bibinfo {volume} {100}},\ \bibinfo {pages}
  {094423} (\bibinfo {year} {2019})}\BibitemShut {NoStop}%
\bibitem [{\citenamefont {Xiao}\ \emph {et~al.}(2010)\citenamefont {Xiao},
  \citenamefont {Bauer}, \citenamefont {Uchida}, \citenamefont {Saitoh},\ and\
  \citenamefont {Maekawa}}]{Xiao_2010}%
  \BibitemOpen
  \bibfield  {author} {\bibinfo {author} {\bibfnamefont {Jiang}\ \bibnamefont
  {Xiao}}, \bibinfo {author} {\bibfnamefont {Gerrit E.~W.}\ \bibnamefont
  {Bauer}}, \bibinfo {author} {\bibfnamefont {Ken-chi}\ \bibnamefont {Uchida}},
  \bibinfo {author} {\bibfnamefont {Eiji}\ \bibnamefont {Saitoh}}, \ and\
  \bibinfo {author} {\bibfnamefont {Sadamichi}\ \bibnamefont {Maekawa}},\
  }\bibfield  {title} {\enquote {\bibinfo {title} {Theory of magnon-driven spin
  seebeck effect},}\ }\href {\doibase 10.1103/PhysRevB.81.214418} {\bibfield
  {journal} {\bibinfo  {journal} {Phys. Rev. B}\ }\textbf {\bibinfo {volume}
  {81}},\ \bibinfo {pages} {214418} (\bibinfo {year} {2010})}\BibitemShut
  {NoStop}%
\bibitem [{\citenamefont {\textit{et al.}}(2014)}]{Rezende_2014}%
  \BibitemOpen
  \bibfield  {author} {\bibinfo {author} {\bibfnamefont {S.~M.~Rezende}\
  \bibnamefont {\textit{et al.}}},\ }\bibfield  {title} {\enquote {\bibinfo
  {title} {Magnon spin-current theory for the longitudinal spin-seebeck
  effect},}\ }\href {\doibase 10.1103/PhysRevB.89.014416} {\bibfield  {journal}
  {\bibinfo  {journal} {Phys. Rev. B}\ }\textbf {\bibinfo {volume} {89}},\
  \bibinfo {pages} {014416} (\bibinfo {year} {2014})}\BibitemShut {NoStop}%
\bibitem [{\citenamefont {Uchida}\ \emph {et~al.}(2010)\citenamefont {Uchida},
  \citenamefont {Adachi}, \citenamefont {Ota}, \citenamefont {Nakayama},
  \citenamefont {Maekawa},\ and\ \citenamefont {Saitoh}}]{Uchida_2010}%
  \BibitemOpen
  \bibfield  {author} {\bibinfo {author} {\bibfnamefont {Ken-ichi}\
  \bibnamefont {Uchida}}, \bibinfo {author} {\bibfnamefont {Hiroto}\
  \bibnamefont {Adachi}}, \bibinfo {author} {\bibfnamefont {Takeru}\
  \bibnamefont {Ota}}, \bibinfo {author} {\bibfnamefont {Hiroyasu}\
  \bibnamefont {Nakayama}}, \bibinfo {author} {\bibfnamefont {Sadamichi}\
  \bibnamefont {Maekawa}}, \ and\ \bibinfo {author} {\bibfnamefont {Eiji}\
  \bibnamefont {Saitoh}},\ }\bibfield  {title} {\enquote {\bibinfo {title}
  {Observation of longitudinal spin-seebeck effect in magnetic insulators},}\
  }\href {https://doi.org/10.1063/1.3507386} {\bibfield  {journal} {\bibinfo
  {journal} {Applied Physics Letters}\ }\textbf {\bibinfo {volume} {97}}
  (\bibinfo {year} {2010})}\BibitemShut {NoStop}%
\bibitem [{\citenamefont {Seshadri}\ and\ \citenamefont
  {Sen}(2018)}]{Seshadri_2018}%
  \BibitemOpen
  \bibfield  {author} {\bibinfo {author} {\bibfnamefont {Ranjani}\ \bibnamefont
  {Seshadri}}\ and\ \bibinfo {author} {\bibfnamefont {Diptiman}\ \bibnamefont
  {Sen}},\ }\bibfield  {title} {\enquote {\bibinfo {title} {Topological magnons
  in a kagome-lattice spin system with $xxz$ and dzyaloshinskii-moriya
  interactions},}\ }\href {\doibase 10.1103/PhysRevB.97.134411} {\bibfield
  {journal} {\bibinfo  {journal} {Phys. Rev. B}\ }\textbf {\bibinfo {volume}
  {97}},\ \bibinfo {pages} {134411} (\bibinfo {year} {2018})}\BibitemShut
  {NoStop}%
\bibitem [{\citenamefont {\ifmmode~\check{S}\else \v{S}\fi{}mejkal}\ \emph
  {et~al.}(2022{\natexlab{a}})\citenamefont {\ifmmode~\check{S}\else
  \v{S}\fi{}mejkal}, \citenamefont {Sinova},\ and\ \citenamefont
  {Jungwirth}}]{Sinova_2022}%
  \BibitemOpen
  \bibfield  {author} {\bibinfo {author} {\bibfnamefont {Libor}\ \bibnamefont
  {\ifmmode~\check{S}\else \v{S}\fi{}mejkal}}, \bibinfo {author} {\bibfnamefont
  {Jairo}\ \bibnamefont {Sinova}}, \ and\ \bibinfo {author} {\bibfnamefont
  {Tomas}\ \bibnamefont {Jungwirth}},\ }\bibfield  {title} {\enquote {\bibinfo
  {title} {Emerging research landscape of altermagnetism},}\ }\href {\doibase
  10.1103/PhysRevX.12.040501} {\bibfield  {journal} {\bibinfo  {journal} {Phys.
  Rev. X}\ }\textbf {\bibinfo {volume} {12}},\ \bibinfo {pages} {040501}
  (\bibinfo {year} {2022}{\natexlab{a}})}\BibitemShut {NoStop}%
\bibitem [{\citenamefont {Fender}\ \emph {et~al.}(2025)\citenamefont {Fender},
  \citenamefont {Gonzalez},\ and\ \citenamefont {Bediako}}]{Fender_2025}%
  \BibitemOpen
  \bibfield  {author} {\bibinfo {author} {\bibfnamefont {Shannon~S.}\
  \bibnamefont {Fender}}, \bibinfo {author} {\bibfnamefont {Oscar}\
  \bibnamefont {Gonzalez}}, \ and\ \bibinfo {author} {\bibfnamefont
  {D.~Kwabena}\ \bibnamefont {Bediako}},\ }\bibfield  {title} {\enquote
  {\bibinfo {title} {Altermagnetism: A chemical perspective},}\ }\href
  {\doibase 10.1021/jacs.4c14503} {\bibfield  {journal} {\bibinfo  {journal}
  {Journal of the American Chemical Society}\ }\textbf {\bibinfo {volume}
  {147}},\ \bibinfo {pages} {2257--2274} (\bibinfo {year} {2025})}\BibitemShut
  {NoStop}%
\bibitem [{\citenamefont {Gomonay}\ \emph {et~al.}(2024)\citenamefont
  {Gomonay}, \citenamefont {Kravchuk}, \citenamefont {Jaeschke-Ubiergo},
  \citenamefont {Yershov}, \citenamefont {Jungwirth}, \citenamefont
  {{\v{S}}mejkal}, \citenamefont {Brink},\ and\ \citenamefont
  {Sinova}}]{Gomonay_2024}%
  \BibitemOpen
  \bibfield  {author} {\bibinfo {author} {\bibfnamefont {O}~\bibnamefont
  {Gomonay}}, \bibinfo {author} {\bibfnamefont {VP}~\bibnamefont {Kravchuk}},
  \bibinfo {author} {\bibfnamefont {R}~\bibnamefont {Jaeschke-Ubiergo}},
  \bibinfo {author} {\bibfnamefont {KV}~\bibnamefont {Yershov}}, \bibinfo
  {author} {\bibfnamefont {T}~\bibnamefont {Jungwirth}}, \bibinfo {author}
  {\bibfnamefont {L}~\bibnamefont {{\v{S}}mejkal}}, \bibinfo {author}
  {\bibfnamefont {J~van~den}\ \bibnamefont {Brink}}, \ and\ \bibinfo {author}
  {\bibfnamefont {J}~\bibnamefont {Sinova}},\ }\bibfield  {title} {\enquote
  {\bibinfo {title} {Structure, control, and dynamics of altermagnetic
  textures},}\ }\href {https://www.nature.com/articles/s44306-024-00042-3}
  {\bibfield  {journal} {\bibinfo  {journal} {npj Spintronics}\ }\textbf
  {\bibinfo {volume} {2}},\ \bibinfo {pages} {35} (\bibinfo {year}
  {2024})}\BibitemShut {NoStop}%
\bibitem [{\citenamefont {Cui}\ \emph {et~al.}(2023)\citenamefont {Cui},
  \citenamefont {Zeng}, \citenamefont {Cui}, \citenamefont {Yu},\ and\
  \citenamefont {Yang}}]{Cui_2023}%
  \BibitemOpen
  \bibfield  {author} {\bibinfo {author} {\bibfnamefont {Qirui}\ \bibnamefont
  {Cui}}, \bibinfo {author} {\bibfnamefont {Bowen}\ \bibnamefont {Zeng}},
  \bibinfo {author} {\bibfnamefont {Ping}\ \bibnamefont {Cui}}, \bibinfo
  {author} {\bibfnamefont {Tao}\ \bibnamefont {Yu}}, \ and\ \bibinfo {author}
  {\bibfnamefont {Hongxin}\ \bibnamefont {Yang}},\ }\bibfield  {title}
  {\enquote {\bibinfo {title} {Efficient spin seebeck and spin nernst effects
  of magnons in altermagnets},}\ }\href {\doibase 10.1103/PhysRevB.108.L180401}
  {\bibfield  {journal} {\bibinfo  {journal} {Phys. Rev. B}\ }\textbf {\bibinfo
  {volume} {108}},\ \bibinfo {pages} {L180401} (\bibinfo {year}
  {2023})}\BibitemShut {NoStop}%
\bibitem [{\citenamefont {Krempask{\`y}}\ \emph {et~al.}(2024)\citenamefont
  {Krempask{\`y}}, \citenamefont {{\v{S}}mejkal}, \citenamefont {D’souza},
  \citenamefont {Hajlaoui}, \citenamefont {Springholz}, \citenamefont
  {Uhl{\'\i}{\v{r}}ov{\'a}}, \citenamefont {Alarab}, \citenamefont
  {Constantinou}, \citenamefont {Strocov}, \citenamefont {Usanov} \emph
  {et~al.}}]{krempasky_2024}%
  \BibitemOpen
  \bibfield  {author} {\bibinfo {author} {\bibfnamefont {Juraj}\ \bibnamefont
  {Krempask{\`y}}}, \bibinfo {author} {\bibfnamefont {L}~\bibnamefont
  {{\v{S}}mejkal}}, \bibinfo {author} {\bibfnamefont {SW}~\bibnamefont
  {D’souza}}, \bibinfo {author} {\bibfnamefont {M}~\bibnamefont {Hajlaoui}},
  \bibinfo {author} {\bibfnamefont {G}~\bibnamefont {Springholz}}, \bibinfo
  {author} {\bibfnamefont {K}~\bibnamefont {Uhl{\'\i}{\v{r}}ov{\'a}}}, \bibinfo
  {author} {\bibfnamefont {F}~\bibnamefont {Alarab}}, \bibinfo {author}
  {\bibfnamefont {PC}~\bibnamefont {Constantinou}}, \bibinfo {author}
  {\bibfnamefont {V}~\bibnamefont {Strocov}}, \bibinfo {author} {\bibfnamefont
  {D}~\bibnamefont {Usanov}},  \emph {et~al.},\ }\bibfield  {title} {\enquote
  {\bibinfo {title} {Altermagnetic lifting of kramers spin degeneracy},}\
  }\href {https://www.nature.com/articles/s41586-023-06907-7} {\bibfield
  {journal} {\bibinfo  {journal} {Nature}\ }\textbf {\bibinfo {volume} {626}},\
  \bibinfo {pages} {517--522} (\bibinfo {year} {2024})}\BibitemShut {NoStop}%
\bibitem [{\citenamefont {Liu}\ \emph {et~al.}(2024{\natexlab{a}})\citenamefont
  {Liu}, \citenamefont {Ozeki}, \citenamefont {Asai}, \citenamefont {Itoh},\
  and\ \citenamefont {Masuda}}]{Liu_2024}%
  \BibitemOpen
  \bibfield  {author} {\bibinfo {author} {\bibfnamefont {Zheyuan}\ \bibnamefont
  {Liu}}, \bibinfo {author} {\bibfnamefont {Makoto}\ \bibnamefont {Ozeki}},
  \bibinfo {author} {\bibfnamefont {Shinichiro}\ \bibnamefont {Asai}}, \bibinfo
  {author} {\bibfnamefont {Shinichi}\ \bibnamefont {Itoh}}, \ and\ \bibinfo
  {author} {\bibfnamefont {Takatsugu}\ \bibnamefont {Masuda}},\ }\bibfield
  {title} {\enquote {\bibinfo {title} {Chiral split magnon in altermagnetic
  mnte},}\ }\href {\doibase 10.1103/PhysRevLett.133.156702} {\bibfield
  {journal} {\bibinfo  {journal} {Phys. Rev. Lett.}\ }\textbf {\bibinfo
  {volume} {133}},\ \bibinfo {pages} {156702} (\bibinfo {year}
  {2024}{\natexlab{a}})}\BibitemShut {NoStop}%
\bibitem [{\citenamefont {\ifmmode~\check{S}\else \v{S}\fi{}mejkal}\ \emph
  {et~al.}(2022{\natexlab{b}})\citenamefont {\ifmmode~\check{S}\else
  \v{S}\fi{}mejkal}, \citenamefont {Sinova},\ and\ \citenamefont
  {Jungwirth}}]{Jairo_2022}%
  \BibitemOpen
  \bibfield  {author} {\bibinfo {author} {\bibfnamefont {Libor}\ \bibnamefont
  {\ifmmode~\check{S}\else \v{S}\fi{}mejkal}}, \bibinfo {author} {\bibfnamefont
  {Jairo}\ \bibnamefont {Sinova}}, \ and\ \bibinfo {author} {\bibfnamefont
  {Tomas}\ \bibnamefont {Jungwirth}},\ }\bibfield  {title} {\enquote {\bibinfo
  {title} {Beyond conventional ferromagnetism and antiferromagnetism: A phase
  with nonrelativistic spin and crystal rotation symmetry},}\ }\href {\doibase
  10.1103/PhysRevX.12.031042} {\bibfield  {journal} {\bibinfo  {journal} {Phys.
  Rev. X}\ }\textbf {\bibinfo {volume} {12}},\ \bibinfo {pages} {031042}
  (\bibinfo {year} {2022}{\natexlab{b}})}\BibitemShut {NoStop}%
\bibitem [{\citenamefont {Brekke}\ \emph {et~al.}(2023)\citenamefont {Brekke},
  \citenamefont {Brataas},\ and\ \citenamefont {Sudb\o{}}}]{Brekke_2023}%
  \BibitemOpen
  \bibfield  {author} {\bibinfo {author} {\bibfnamefont {Bj\o{}rnulf}\
  \bibnamefont {Brekke}}, \bibinfo {author} {\bibfnamefont {Arne}\ \bibnamefont
  {Brataas}}, \ and\ \bibinfo {author} {\bibfnamefont {Asle}\ \bibnamefont
  {Sudb\o{}}},\ }\bibfield  {title} {\enquote {\bibinfo {title}
  {Two-dimensional altermagnets: Superconductivity in a minimal microscopic
  model},}\ }\href {\doibase 10.1103/PhysRevB.108.224421} {\bibfield  {journal}
  {\bibinfo  {journal} {Phys. Rev. B}\ }\textbf {\bibinfo {volume} {108}},\
  \bibinfo {pages} {224421} (\bibinfo {year} {2023})}\BibitemShut {NoStop}%
\bibitem [{\citenamefont {Wei\ss{}enhofer}\ and\ \citenamefont
  {Marmodoro}(2024)}]{Alberto_2024}%
  \BibitemOpen
  \bibfield  {author} {\bibinfo {author} {\bibfnamefont {Markus}\ \bibnamefont
  {Wei\ss{}enhofer}}\ and\ \bibinfo {author} {\bibfnamefont {Alberto}\
  \bibnamefont {Marmodoro}},\ }\bibfield  {title} {\enquote {\bibinfo {title}
  {Atomistic spin dynamics simulations of magnonic spin seebeck and spin nernst
  effects in altermagnets},}\ }\href {\doibase 10.1103/PhysRevB.110.094427}
  {\bibfield  {journal} {\bibinfo  {journal} {Phys. Rev. B}\ }\textbf {\bibinfo
  {volume} {110}},\ \bibinfo {pages} {094427} (\bibinfo {year}
  {2024})}\BibitemShut {NoStop}%
\bibitem [{\citenamefont {\textit{et al.}}(2024{\natexlab{a}})}]{Lee_2024}%
  \BibitemOpen
  \bibfield  {author} {\bibinfo {author} {\bibfnamefont {Suyoung~Lee}\
  \bibnamefont {\textit{et al.}}},\ }\bibfield  {title} {\enquote {\bibinfo
  {title} {Broken kramers degeneracy in altermagnetic mnte},}\ }\href {\doibase
  10.1103/PhysRevLett.132.036702} {\bibfield  {journal} {\bibinfo  {journal}
  {Phys. Rev. Lett.}\ }\textbf {\bibinfo {volume} {132}},\ \bibinfo {pages}
  {036702} (\bibinfo {year} {2024}{\natexlab{a}})}\BibitemShut {NoStop}%
\bibitem [{\citenamefont {Camerano}\ \emph
  {et~al.}(2025{\natexlab{a}})\citenamefont {Camerano}, \citenamefont {Fumega},
  \citenamefont {Profeta},\ and\ \citenamefont {Lado}}]{Camerano_2025}%
  \BibitemOpen
  \bibfield  {author} {\bibinfo {author} {\bibfnamefont {Luigi}\ \bibnamefont
  {Camerano}}, \bibinfo {author} {\bibfnamefont {Adolfo~O.}\ \bibnamefont
  {Fumega}}, \bibinfo {author} {\bibfnamefont {Gianni}\ \bibnamefont
  {Profeta}}, \ and\ \bibinfo {author} {\bibfnamefont {Jose~L.}\ \bibnamefont
  {Lado}},\ }\bibfield  {title} {\enquote {\bibinfo {title} {Multicomponent
  magneto-orbital order and magneto-orbitons in monolayer vcl3},}\ }\href
  {\doibase 10.1021/acs.nanolett.4c06400} {\bibfield  {journal} {\bibinfo
  {journal} {Nano Letters}\ }\textbf {\bibinfo {volume} {25}},\ \bibinfo
  {pages} {4825--4831} (\bibinfo {year} {2025}{\natexlab{a}})},\ \bibinfo
  {note} {pMID: 39960815}\BibitemShut {NoStop}%
\bibitem [{\citenamefont {Camerano}\ \emph
  {et~al.}(2025{\natexlab{b}})\citenamefont {Camerano}, \citenamefont {Fumega},
  \citenamefont {Lado}, \citenamefont {Stroppa},\ and\ \citenamefont
  {Profeta}}]{camerano_2025_2}%
  \BibitemOpen
  \bibfield  {author} {\bibinfo {author} {\bibfnamefont {Luigi}\ \bibnamefont
  {Camerano}}, \bibinfo {author} {\bibfnamefont {Adolfo~O.}\ \bibnamefont
  {Fumega}}, \bibinfo {author} {\bibfnamefont {Jose~L.}\ \bibnamefont {Lado}},
  \bibinfo {author} {\bibfnamefont {Alessandro}\ \bibnamefont {Stroppa}}, \
  and\ \bibinfo {author} {\bibfnamefont {Gianni}\ \bibnamefont {Profeta}},\
  }\href {https://arxiv.org/abs/2503.19987} {\enquote {\bibinfo {title}
  {Multiferroic nematic d-wave altermagnetism driven by orbital-order on the
  honeycomb lattice},}\ } (\bibinfo {year} {2025}{\natexlab{b}}),\ \Eprint
  {http://arxiv.org/abs/2503.19987} {arXiv:2503.19987 [cond-mat.mtrl-sci]}
  \BibitemShut {NoStop}%
\bibitem [{\citenamefont {Varshney}\ \emph
  {et~al.}(2023{\natexlab{a}})\citenamefont {Varshney}, \citenamefont
  {Mukherjee}, \citenamefont {Kundu},\ and\ \citenamefont
  {Agarwal}}]{Harsh_2023}%
  \BibitemOpen
  \bibfield  {author} {\bibinfo {author} {\bibfnamefont {Harsh}\ \bibnamefont
  {Varshney}}, \bibinfo {author} {\bibfnamefont {Rohit}\ \bibnamefont
  {Mukherjee}}, \bibinfo {author} {\bibfnamefont {Arijit}\ \bibnamefont
  {Kundu}}, \ and\ \bibinfo {author} {\bibfnamefont {Amit}\ \bibnamefont
  {Agarwal}},\ }\bibfield  {title} {\enquote {\bibinfo {title} {Intrinsic
  nonlinear thermal hall transport of magnons: A quantum kinetic theory
  approach},}\ }\href {\doibase 10.1103/PhysRevB.108.165412} {\bibfield
  {journal} {\bibinfo  {journal} {Phys. Rev. B}\ }\textbf {\bibinfo {volume}
  {108}},\ \bibinfo {pages} {165412} (\bibinfo {year}
  {2023}{\natexlab{a}})}\BibitemShut {NoStop}%
\bibitem [{\citenamefont {Varshney}\ \emph
  {et~al.}(2023{\natexlab{b}})\citenamefont {Varshney}, \citenamefont {Das},
  \citenamefont {Bhalla},\ and\ \citenamefont {Agarwal}}]{Harsh_2023_2}%
  \BibitemOpen
  \bibfield  {author} {\bibinfo {author} {\bibfnamefont {Harsh}\ \bibnamefont
  {Varshney}}, \bibinfo {author} {\bibfnamefont {Kamal}\ \bibnamefont {Das}},
  \bibinfo {author} {\bibfnamefont {Pankaj}\ \bibnamefont {Bhalla}}, \ and\
  \bibinfo {author} {\bibfnamefont {Amit}\ \bibnamefont {Agarwal}},\ }\bibfield
   {title} {\enquote {\bibinfo {title} {Quantum kinetic theory of nonlinear
  thermal current},}\ }\href {\doibase 10.1103/PhysRevB.107.235419} {\bibfield
  {journal} {\bibinfo  {journal} {Phys. Rev. B}\ }\textbf {\bibinfo {volume}
  {107}},\ \bibinfo {pages} {235419} (\bibinfo {year}
  {2023}{\natexlab{b}})}\BibitemShut {NoStop}%
\bibitem [{\citenamefont {Varshney}\ and\ \citenamefont
  {Agarwal}(2024)}]{Harsh_2024}%
  \BibitemOpen
  \bibfield  {author} {\bibinfo {author} {\bibfnamefont {Harsh}\ \bibnamefont
  {Varshney}}\ and\ \bibinfo {author} {\bibfnamefont {Amit}\ \bibnamefont
  {Agarwal}},\ }\bibfield  {title} {\enquote {\bibinfo {title} {Intrinsic
  nonlinear nernst and seebeck effect},}\ }\href
  {https://arxiv.org/abs/2409.11108} {\bibfield  {journal} {\bibinfo  {journal}
  {arXiv preprint arXiv:2409.11108}\ } (\bibinfo {year} {2024})}\BibitemShut
  {NoStop}%
\bibitem [{\citenamefont {Sarkar}\ \emph {et~al.}(2025)\citenamefont {Sarkar},
  \citenamefont {Das},\ and\ \citenamefont {Agarwal}}]{Sayan_2025}%
  \BibitemOpen
  \bibfield  {author} {\bibinfo {author} {\bibfnamefont {Sayan}\ \bibnamefont
  {Sarkar}}, \bibinfo {author} {\bibfnamefont {Sunit}\ \bibnamefont {Das}}, \
  and\ \bibinfo {author} {\bibfnamefont {Amit}\ \bibnamefont {Agarwal}},\
  }\bibfield  {title} {\enquote {\bibinfo {title} {Symmetry-driven intrinsic
  nonlinear pure spin hall effect},}\ \ }(\bibinfo {year} {2025})\BibitemShut
  {NoStop}%
\bibitem [{\citenamefont {Sarkar}\ \emph {et~al.}(2024)\citenamefont {Sarkar},
  \citenamefont {Das}, \citenamefont {Mandal},\ and\ \citenamefont
  {Agarwal}}]{Sayan_2024}%
  \BibitemOpen
  \bibfield  {author} {\bibinfo {author} {\bibfnamefont {Sayan}\ \bibnamefont
  {Sarkar}}, \bibinfo {author} {\bibfnamefont {Sunit}\ \bibnamefont {Das}},
  \bibinfo {author} {\bibfnamefont {Debottam}\ \bibnamefont {Mandal}}, \ and\
  \bibinfo {author} {\bibfnamefont {Amit}\ \bibnamefont {Agarwal}},\ }\bibfield
   {title} {\enquote {\bibinfo {title} {Light-induced nonlinear resonant spin
  magnetization},}\ }\href {https://arxiv.org/abs/2409.12142} {\bibfield
  {journal} {\bibinfo  {journal} {arXiv preprint arXiv:2409.12142}\ } (\bibinfo
  {year} {2024})}\BibitemShut {NoStop}%
\bibitem [{\citenamefont {Sekine}\ and\ \citenamefont
  {Nagaosa}(2020)}]{Sekine_2020}%
  \BibitemOpen
  \bibfield  {author} {\bibinfo {author} {\bibfnamefont {Akihiko}\ \bibnamefont
  {Sekine}}\ and\ \bibinfo {author} {\bibfnamefont {Naoto}\ \bibnamefont
  {Nagaosa}},\ }\bibfield  {title} {\enquote {\bibinfo {title} {Quantum kinetic
  theory of thermoelectric and thermal transport in a magnetic field},}\ }\href
  {\doibase 10.1103/PhysRevB.101.155204} {\bibfield  {journal} {\bibinfo
  {journal} {Phys. Rev. B}\ }\textbf {\bibinfo {volume} {101}},\ \bibinfo
  {pages} {155204} (\bibinfo {year} {2020})}\BibitemShut {NoStop}%
\bibitem [{\citenamefont {Culcer}\ \emph {et~al.}(2017)\citenamefont {Culcer},
  \citenamefont {Sekine},\ and\ \citenamefont {MacDonald}}]{Culcer_2017}%
  \BibitemOpen
  \bibfield  {author} {\bibinfo {author} {\bibfnamefont {Dimitrie}\
  \bibnamefont {Culcer}}, \bibinfo {author} {\bibfnamefont {Akihiko}\
  \bibnamefont {Sekine}}, \ and\ \bibinfo {author} {\bibfnamefont {Allan~H.}\
  \bibnamefont {MacDonald}},\ }\bibfield  {title} {\enquote {\bibinfo {title}
  {Interband coherence response to electric fields in crystals: Berry-phase
  contributions and disorder effects},}\ }\href {\doibase
  10.1103/PhysRevB.96.035106} {\bibfield  {journal} {\bibinfo  {journal} {Phys.
  Rev. B}\ }\textbf {\bibinfo {volume} {96}},\ \bibinfo {pages} {035106}
  (\bibinfo {year} {2017})}\BibitemShut {NoStop}%
\bibitem [{\citenamefont {Boyd}\ \emph {et~al.}(2008)\citenamefont {Boyd},
  \citenamefont {Gaeta},\ and\ \citenamefont {Giese}}]{boyd_2008}%
  \BibitemOpen
  \bibfield  {author} {\bibinfo {author} {\bibfnamefont {Robert~W}\
  \bibnamefont {Boyd}}, \bibinfo {author} {\bibfnamefont {Alexander~L}\
  \bibnamefont {Gaeta}}, \ and\ \bibinfo {author} {\bibfnamefont {Enno}\
  \bibnamefont {Giese}},\ }\bibfield  {title} {\enquote {\bibinfo {title}
  {Nonlinear optics},}\ }in\ \href@noop {} {\emph {\bibinfo {booktitle}
  {Springer Handbook of Atomic, Molecular, and Optical Physics}}}\ (\bibinfo
  {publisher} {Springer},\ \bibinfo {year} {2008})\ pp.\ \bibinfo {pages}
  {1097--1110}\BibitemShut {NoStop}%
\bibitem [{\citenamefont {Tatara}(2015)}]{Tatara_2015}%
  \BibitemOpen
  \bibfield  {author} {\bibinfo {author} {\bibfnamefont {Gen}\ \bibnamefont
  {Tatara}},\ }\bibfield  {title} {\enquote {\bibinfo {title} {Thermal vector
  potential theory of transport induced by a temperature gradient},}\ }\href
  {\doibase 10.1103/PhysRevLett.114.196601} {\bibfield  {journal} {\bibinfo
  {journal} {Phys. Rev. Lett.}\ }\textbf {\bibinfo {volume} {114}},\ \bibinfo
  {pages} {196601} (\bibinfo {year} {2015})}\BibitemShut {NoStop}%
\bibitem [{\citenamefont {Gao}\ \emph {et~al.}(2021)\citenamefont {Gao},
  \citenamefont {Liu}, \citenamefont {Hu}, \citenamefont {Qiu}, \citenamefont
  {Tzschaschel}, \citenamefont {Ghosh}, \citenamefont {Ho}, \citenamefont
  {B{\'e}rub{\'e}}, \citenamefont {Chen}, \citenamefont {Sun} \emph
  {et~al.}}]{Gao_2021}%
  \BibitemOpen
  \bibfield  {author} {\bibinfo {author} {\bibfnamefont {Anyuan}\ \bibnamefont
  {Gao}}, \bibinfo {author} {\bibfnamefont {Yu-Fei}\ \bibnamefont {Liu}},
  \bibinfo {author} {\bibfnamefont {Chaowei}\ \bibnamefont {Hu}}, \bibinfo
  {author} {\bibfnamefont {Jian-Xiang}\ \bibnamefont {Qiu}}, \bibinfo {author}
  {\bibfnamefont {Christian}\ \bibnamefont {Tzschaschel}}, \bibinfo {author}
  {\bibfnamefont {Barun}\ \bibnamefont {Ghosh}}, \bibinfo {author}
  {\bibfnamefont {Sheng-Chin}\ \bibnamefont {Ho}}, \bibinfo {author}
  {\bibfnamefont {Damien}\ \bibnamefont {B{\'e}rub{\'e}}}, \bibinfo {author}
  {\bibfnamefont {Rui}\ \bibnamefont {Chen}}, \bibinfo {author} {\bibfnamefont
  {Haipeng}\ \bibnamefont {Sun}},  \emph {et~al.},\ }\bibfield  {title}
  {\enquote {\bibinfo {title} {Layer hall effect in a 2d topological axion
  antiferromagnet},}\ }\href
  {https://www.nature.com/articles/s41586-021-03679-w} {\bibfield  {journal}
  {\bibinfo  {journal} {Nature}\ }\textbf {\bibinfo {volume} {595}},\ \bibinfo
  {pages} {521--525} (\bibinfo {year} {2021})}\BibitemShut {NoStop}%
\bibitem [{\citenamefont {Sinha}\ \emph {et~al.}(2022)\citenamefont {Sinha},
  \citenamefont {Adak}, \citenamefont {Chakraborty}, \citenamefont {Das},
  \citenamefont {Debnath}, \citenamefont {Sangani}, \citenamefont {Watanabe},
  \citenamefont {Taniguchi}, \citenamefont {Waghmare}, \citenamefont {Agarwal}
  \emph {et~al.}}]{Sinha_2022}%
  \BibitemOpen
  \bibfield  {author} {\bibinfo {author} {\bibfnamefont {Subhajit}\
  \bibnamefont {Sinha}}, \bibinfo {author} {\bibfnamefont {Pratap~Chandra}\
  \bibnamefont {Adak}}, \bibinfo {author} {\bibfnamefont {Atasi}\ \bibnamefont
  {Chakraborty}}, \bibinfo {author} {\bibfnamefont {Kamal}\ \bibnamefont
  {Das}}, \bibinfo {author} {\bibfnamefont {Koyendrila}\ \bibnamefont
  {Debnath}}, \bibinfo {author} {\bibfnamefont {LD~Varma}\ \bibnamefont
  {Sangani}}, \bibinfo {author} {\bibfnamefont {Kenji}\ \bibnamefont
  {Watanabe}}, \bibinfo {author} {\bibfnamefont {Takashi}\ \bibnamefont
  {Taniguchi}}, \bibinfo {author} {\bibfnamefont {Umesh~V}\ \bibnamefont
  {Waghmare}}, \bibinfo {author} {\bibfnamefont {Amit}\ \bibnamefont
  {Agarwal}},  \emph {et~al.},\ }\bibfield  {title} {\enquote {\bibinfo {title}
  {Berry curvature dipole senses topological transition in a moir{\'e}
  superlattice},}\ }\href {https://www.nature.com/articles/s41567-022-01606-y}
  {\bibfield  {journal} {\bibinfo  {journal} {Nature Physics}\ }\textbf
  {\bibinfo {volume} {18}},\ \bibinfo {pages} {765--770} (\bibinfo {year}
  {2022})}\BibitemShut {NoStop}%
\bibitem [{\citenamefont {Mandal}\ \emph {et~al.}(2024)\citenamefont {Mandal},
  \citenamefont {Sarkar}, \citenamefont {Das},\ and\ \citenamefont
  {Agarwal}}]{Debottam_2024}%
  \BibitemOpen
  \bibfield  {author} {\bibinfo {author} {\bibfnamefont {Debottam}\
  \bibnamefont {Mandal}}, \bibinfo {author} {\bibfnamefont {Sanjay}\
  \bibnamefont {Sarkar}}, \bibinfo {author} {\bibfnamefont {Kamal}\
  \bibnamefont {Das}}, \ and\ \bibinfo {author} {\bibfnamefont {Amit}\
  \bibnamefont {Agarwal}},\ }\bibfield  {title} {\enquote {\bibinfo {title}
  {Quantum geometry induced third-order nonlinear transport responses},}\
  }\href {\doibase 10.1103/PhysRevB.110.195131} {\bibfield  {journal} {\bibinfo
   {journal} {Phys. Rev. B}\ }\textbf {\bibinfo {volume} {110}},\ \bibinfo
  {pages} {195131} (\bibinfo {year} {2024})}\BibitemShut {NoStop}%
\bibitem [{\citenamefont {Das}\ \emph {et~al.}(2023)\citenamefont {Das},
  \citenamefont {Lahiri}, \citenamefont {Atencia}, \citenamefont {Culcer},\
  and\ \citenamefont {Agarwal}}]{Kamal_2023}%
  \BibitemOpen
  \bibfield  {author} {\bibinfo {author} {\bibfnamefont {Kamal}\ \bibnamefont
  {Das}}, \bibinfo {author} {\bibfnamefont {Shibalik}\ \bibnamefont {Lahiri}},
  \bibinfo {author} {\bibfnamefont {Rhonald~Burgos}\ \bibnamefont {Atencia}},
  \bibinfo {author} {\bibfnamefont {Dimitrie}\ \bibnamefont {Culcer}}, \ and\
  \bibinfo {author} {\bibfnamefont {Amit}\ \bibnamefont {Agarwal}},\ }\bibfield
   {title} {\enquote {\bibinfo {title} {Intrinsic nonlinear conductivities
  induced by the quantum metric},}\ }\href {\doibase
  10.1103/PhysRevB.108.L201405} {\bibfield  {journal} {\bibinfo  {journal}
  {Phys. Rev. B}\ }\textbf {\bibinfo {volume} {108}},\ \bibinfo {pages}
  {L201405} (\bibinfo {year} {2023})}\BibitemShut {NoStop}%
\bibitem [{\citenamefont {Seki}\ \emph {et~al.}(2015)\citenamefont {Seki},
  \citenamefont {Ideue}, \citenamefont {Kubota}, \citenamefont {Kozuka},
  \citenamefont {Takagi}, \citenamefont {Nakamura}, \citenamefont {Kaneko},
  \citenamefont {Kawasaki},\ and\ \citenamefont {Tokura}}]{Seki_2015}%
  \BibitemOpen
  \bibfield  {author} {\bibinfo {author} {\bibfnamefont {S.}~\bibnamefont
  {Seki}}, \bibinfo {author} {\bibfnamefont {T.}~\bibnamefont {Ideue}},
  \bibinfo {author} {\bibfnamefont {M.}~\bibnamefont {Kubota}}, \bibinfo
  {author} {\bibfnamefont {Y.}~\bibnamefont {Kozuka}}, \bibinfo {author}
  {\bibfnamefont {R.}~\bibnamefont {Takagi}}, \bibinfo {author} {\bibfnamefont
  {M.}~\bibnamefont {Nakamura}}, \bibinfo {author} {\bibfnamefont
  {Y.}~\bibnamefont {Kaneko}}, \bibinfo {author} {\bibfnamefont
  {M.}~\bibnamefont {Kawasaki}}, \ and\ \bibinfo {author} {\bibfnamefont
  {Y.}~\bibnamefont {Tokura}},\ }\bibfield  {title} {\enquote {\bibinfo {title}
  {Thermal generation of spin current in an antiferromagnet},}\ }\href
  {\doibase 10.1103/PhysRevLett.115.266601} {\bibfield  {journal} {\bibinfo
  {journal} {Phys. Rev. Lett.}\ }\textbf {\bibinfo {volume} {115}},\ \bibinfo
  {pages} {266601} (\bibinfo {year} {2015})}\BibitemShut {NoStop}%
\bibitem [{\citenamefont {Matsumoto}\ and\ \citenamefont
  {Murakami}(2011{\natexlab{a}})}]{Matsumoto_2011}%
  \BibitemOpen
  \bibfield  {author} {\bibinfo {author} {\bibfnamefont {Ryo}\ \bibnamefont
  {Matsumoto}}\ and\ \bibinfo {author} {\bibfnamefont {Shuichi}\ \bibnamefont
  {Murakami}},\ }\bibfield  {title} {\enquote {\bibinfo {title} {Rotational
  motion of magnons and the thermal hall effect},}\ }\href {\doibase
  10.1103/PhysRevB.84.184406} {\bibfield  {journal} {\bibinfo  {journal} {Phys.
  Rev. B}\ }\textbf {\bibinfo {volume} {84}},\ \bibinfo {pages} {184406}
  (\bibinfo {year} {2011}{\natexlab{a}})}\BibitemShut {NoStop}%
\bibitem [{\citenamefont {Matsumoto}\ and\ \citenamefont
  {Murakami}(2011{\natexlab{b}})}]{Matsumoto_2011_prl}%
  \BibitemOpen
  \bibfield  {author} {\bibinfo {author} {\bibfnamefont {Ryo}\ \bibnamefont
  {Matsumoto}}\ and\ \bibinfo {author} {\bibfnamefont {Shuichi}\ \bibnamefont
  {Murakami}},\ }\bibfield  {title} {\enquote {\bibinfo {title} {Theoretical
  prediction of a rotating magnon wave packet in ferromagnets},}\ }\href
  {\doibase 10.1103/PhysRevLett.106.197202} {\bibfield  {journal} {\bibinfo
  {journal} {Phys. Rev. Lett.}\ }\textbf {\bibinfo {volume} {106}},\ \bibinfo
  {pages} {197202} (\bibinfo {year} {2011}{\natexlab{b}})}\BibitemShut
  {NoStop}%
\bibitem [{\citenamefont {Adachi}\ \emph {et~al.}(2013)\citenamefont {Adachi},
  \citenamefont {Uchida}, \citenamefont {Saitoh},\ and\ \citenamefont
  {Maekawa}}]{Adachi_2013}%
  \BibitemOpen
  \bibfield  {author} {\bibinfo {author} {\bibfnamefont {Hiroto}\ \bibnamefont
  {Adachi}}, \bibinfo {author} {\bibfnamefont {Ken-ichi}\ \bibnamefont
  {Uchida}}, \bibinfo {author} {\bibfnamefont {Eiji}\ \bibnamefont {Saitoh}}, \
  and\ \bibinfo {author} {\bibfnamefont {Sadamichi}\ \bibnamefont {Maekawa}},\
  }\bibfield  {title} {\enquote {\bibinfo {title} {Theory of the spin seebeck
  effect},}\ }\href {\doibase 10.1088/0034-4885/76/3/036501} {\bibfield
  {journal} {\bibinfo  {journal} {Reports on Progress in Physics}\ }\textbf
  {\bibinfo {volume} {76}},\ \bibinfo {pages} {036501} (\bibinfo {year}
  {2013})}\BibitemShut {NoStop}%
\bibitem [{\citenamefont {Bhalla}\ \emph {et~al.}(2023)\citenamefont {Bhalla},
  \citenamefont {Das}, \citenamefont {Agarwal},\ and\ \citenamefont
  {Culcer}}]{Pankaj_2023}%
  \BibitemOpen
  \bibfield  {author} {\bibinfo {author} {\bibfnamefont {Pankaj}\ \bibnamefont
  {Bhalla}}, \bibinfo {author} {\bibfnamefont {Kamal}\ \bibnamefont {Das}},
  \bibinfo {author} {\bibfnamefont {Amit}\ \bibnamefont {Agarwal}}, \ and\
  \bibinfo {author} {\bibfnamefont {Dimitrie}\ \bibnamefont {Culcer}},\
  }\bibfield  {title} {\enquote {\bibinfo {title} {Quantum kinetic theory of
  nonlinear optical currents: Finite fermi surface and fermi sea
  contributions},}\ }\href {\doibase 10.1103/PhysRevB.107.165131} {\bibfield
  {journal} {\bibinfo  {journal} {Phys. Rev. B}\ }\textbf {\bibinfo {volume}
  {107}},\ \bibinfo {pages} {165131} (\bibinfo {year} {2023})}\BibitemShut
  {NoStop}%
\bibitem [{\citenamefont {Bhalla}\ \emph {et~al.}(2022)\citenamefont {Bhalla},
  \citenamefont {Das}, \citenamefont {Culcer},\ and\ \citenamefont
  {Agarwal}}]{Pankaj_2022}%
  \BibitemOpen
  \bibfield  {author} {\bibinfo {author} {\bibfnamefont {Pankaj}\ \bibnamefont
  {Bhalla}}, \bibinfo {author} {\bibfnamefont {Kamal}\ \bibnamefont {Das}},
  \bibinfo {author} {\bibfnamefont {Dimitrie}\ \bibnamefont {Culcer}}, \ and\
  \bibinfo {author} {\bibfnamefont {Amit}\ \bibnamefont {Agarwal}},\ }\bibfield
   {title} {\enquote {\bibinfo {title} {Resonant second-harmonic generation as
  a probe of quantum geometry},}\ }\href {\doibase
  10.1103/PhysRevLett.129.227401} {\bibfield  {journal} {\bibinfo  {journal}
  {Phys. Rev. Lett.}\ }\textbf {\bibinfo {volume} {129}},\ \bibinfo {pages}
  {227401} (\bibinfo {year} {2022})}\BibitemShut {NoStop}%
\bibitem [{\citenamefont {Kumar}\ \emph {et~al.}(2024)\citenamefont {Kumar},
  \citenamefont {Sarkar},\ and\ \citenamefont {Agarwal}}]{Maneesh_2024}%
  \BibitemOpen
  \bibfield  {author} {\bibinfo {author} {\bibfnamefont {M.~Maneesh}\
  \bibnamefont {Kumar}}, \bibinfo {author} {\bibfnamefont {Sanjay}\
  \bibnamefont {Sarkar}}, \ and\ \bibinfo {author} {\bibfnamefont {Amit}\
  \bibnamefont {Agarwal}},\ }\bibfield  {title} {\enquote {\bibinfo {title}
  {Band geometry induced electro-optic effect and polarization rotation},}\
  }\href {\doibase 10.1103/PhysRevB.110.125401} {\bibfield  {journal} {\bibinfo
   {journal} {Phys. Rev. B}\ }\textbf {\bibinfo {volume} {110}},\ \bibinfo
  {pages} {125401} (\bibinfo {year} {2024})}\BibitemShut {NoStop}%
\bibitem [{\citenamefont {Lovesey}\ \emph {et~al.}(2023)\citenamefont
  {Lovesey}, \citenamefont {Khalyavin},\ and\ \citenamefont {van~der
  Laan}}]{Lovesey_2023}%
  \BibitemOpen
  \bibfield  {author} {\bibinfo {author} {\bibfnamefont {S.~W.}\ \bibnamefont
  {Lovesey}}, \bibinfo {author} {\bibfnamefont {D.~D.}\ \bibnamefont
  {Khalyavin}}, \ and\ \bibinfo {author} {\bibfnamefont {G.}~\bibnamefont
  {van~der Laan}},\ }\bibfield  {title} {\enquote {\bibinfo {title} {Templates
  for magnetic symmetry and altermagnetism in hexagonal mnte},}\ }\href
  {\doibase 10.1103/PhysRevB.108.174437} {\bibfield  {journal} {\bibinfo
  {journal} {Phys. Rev. B}\ }\textbf {\bibinfo {volume} {108}},\ \bibinfo
  {pages} {174437} (\bibinfo {year} {2023})}\BibitemShut {NoStop}%
\bibitem [{\citenamefont {Kubo}(1952)}]{Kubo_1952}%
  \BibitemOpen
  \bibfield  {author} {\bibinfo {author} {\bibfnamefont {Ryogo}\ \bibnamefont
  {Kubo}},\ }\bibfield  {title} {\enquote {\bibinfo {title} {The spin-wave
  theory of antiferromagnetics},}\ }\href {\doibase 10.1103/PhysRev.87.568}
  {\bibfield  {journal} {\bibinfo  {journal} {Phys. Rev.}\ }\textbf {\bibinfo
  {volume} {87}},\ \bibinfo {pages} {568--580} (\bibinfo {year}
  {1952})}\BibitemShut {NoStop}%
\bibitem [{\citenamefont {Karube}\ \emph {et~al.}(2022)\citenamefont {Karube},
  \citenamefont {Tanaka}, \citenamefont {Sugawara}, \citenamefont {Kadoguchi},
  \citenamefont {Kohda},\ and\ \citenamefont {Nitta}}]{Karube_2022}%
  \BibitemOpen
  \bibfield  {author} {\bibinfo {author} {\bibfnamefont {Shutaro}\ \bibnamefont
  {Karube}}, \bibinfo {author} {\bibfnamefont {Takahiro}\ \bibnamefont
  {Tanaka}}, \bibinfo {author} {\bibfnamefont {Daichi}\ \bibnamefont
  {Sugawara}}, \bibinfo {author} {\bibfnamefont {Naohiro}\ \bibnamefont
  {Kadoguchi}}, \bibinfo {author} {\bibfnamefont {Makoto}\ \bibnamefont
  {Kohda}}, \ and\ \bibinfo {author} {\bibfnamefont {Junsaku}\ \bibnamefont
  {Nitta}},\ }\bibfield  {title} {\enquote {\bibinfo {title} {Observation of
  spin-splitter torque in collinear antiferromagnetic ${\mathrm{ruo}}_{2}$},}\
  }\href {\doibase 10.1103/PhysRevLett.129.137201} {\bibfield  {journal}
  {\bibinfo  {journal} {Phys. Rev. Lett.}\ }\textbf {\bibinfo {volume} {129}},\
  \bibinfo {pages} {137201} (\bibinfo {year} {2022})}\BibitemShut {NoStop}%
\bibitem [{\citenamefont {Bai}\ \emph {et~al.}(2022)\citenamefont {Bai},
  \citenamefont {Han}, \citenamefont {Feng}, \citenamefont {Zhou},
  \citenamefont {Su}, \citenamefont {Wang}, \citenamefont {Liao}, \citenamefont
  {Zhu}, \citenamefont {Chen}, \citenamefont {Pan}, \citenamefont {Fan},\ and\
  \citenamefont {Song}}]{Bai_2022}%
  \BibitemOpen
  \bibfield  {author} {\bibinfo {author} {\bibfnamefont {H.}~\bibnamefont
  {Bai}}, \bibinfo {author} {\bibfnamefont {L.}~\bibnamefont {Han}}, \bibinfo
  {author} {\bibfnamefont {X.~Y.}\ \bibnamefont {Feng}}, \bibinfo {author}
  {\bibfnamefont {Y.~J.}\ \bibnamefont {Zhou}}, \bibinfo {author}
  {\bibfnamefont {R.~X.}\ \bibnamefont {Su}}, \bibinfo {author} {\bibfnamefont
  {Q.}~\bibnamefont {Wang}}, \bibinfo {author} {\bibfnamefont {L.~Y.}\
  \bibnamefont {Liao}}, \bibinfo {author} {\bibfnamefont {W.~X.}\ \bibnamefont
  {Zhu}}, \bibinfo {author} {\bibfnamefont {X.~Z.}\ \bibnamefont {Chen}},
  \bibinfo {author} {\bibfnamefont {F.}~\bibnamefont {Pan}}, \bibinfo {author}
  {\bibfnamefont {X.~L.}\ \bibnamefont {Fan}}, \ and\ \bibinfo {author}
  {\bibfnamefont {C.}~\bibnamefont {Song}},\ }\bibfield  {title} {\enquote
  {\bibinfo {title} {Observation of spin splitting torque in a collinear
  antiferromagnet ${\mathrm{ruo}}_{2}$},}\ }\href {\doibase
  10.1103/PhysRevLett.128.197202} {\bibfield  {journal} {\bibinfo  {journal}
  {Phys. Rev. Lett.}\ }\textbf {\bibinfo {volume} {128}},\ \bibinfo {pages}
  {197202} (\bibinfo {year} {2022})}\BibitemShut {NoStop}%
\bibitem [{\citenamefont {Cheng}\ \emph {et~al.}(2018)\citenamefont {Cheng},
  \citenamefont {Xiao},\ and\ \citenamefont {Zhu}}]{Cheng_2018}%
  \BibitemOpen
  \bibfield  {author} {\bibinfo {author} {\bibfnamefont {Ran}\ \bibnamefont
  {Cheng}}, \bibinfo {author} {\bibfnamefont {Di}~\bibnamefont {Xiao}}, \ and\
  \bibinfo {author} {\bibfnamefont {Jian-Gang}\ \bibnamefont {Zhu}},\
  }\bibfield  {title} {\enquote {\bibinfo {title} {Antiferromagnet-based
  magnonic spin-transfer torque},}\ }\href {\doibase
  10.1103/PhysRevB.98.020408} {\bibfield  {journal} {\bibinfo  {journal} {Phys.
  Rev. B}\ }\textbf {\bibinfo {volume} {98}},\ \bibinfo {pages} {020408}
  (\bibinfo {year} {2018})}\BibitemShut {NoStop}%
\bibitem [{\citenamefont {Gonz\'alez-Hern\'andez}\ \emph
  {et~al.}(2021)\citenamefont {Gonz\'alez-Hern\'andez}, \citenamefont
  {\ifmmode~\check{S}\else \v{S}\fi{}mejkal}, \citenamefont {V\'yborn\'y},
  \citenamefont {Yahagi}, \citenamefont {Sinova}, \citenamefont {Jungwirth},\
  and\ \citenamefont {\ifmmode~\check{Z}\else
  \v{Z}\fi{}elezn\'y}}]{Jakub_2021}%
  \BibitemOpen
  \bibfield  {author} {\bibinfo {author} {\bibfnamefont {Rafael}\ \bibnamefont
  {Gonz\'alez-Hern\'andez}}, \bibinfo {author} {\bibfnamefont {Libor}\
  \bibnamefont {\ifmmode~\check{S}\else \v{S}\fi{}mejkal}}, \bibinfo {author}
  {\bibfnamefont {Karel}\ \bibnamefont {V\'yborn\'y}}, \bibinfo {author}
  {\bibfnamefont {Yuta}\ \bibnamefont {Yahagi}}, \bibinfo {author}
  {\bibfnamefont {Jairo}\ \bibnamefont {Sinova}}, \bibinfo {author}
  {\bibfnamefont {Tom\'a\ifmmode \check{s}\else~\v{s}\fi{}}\ \bibnamefont
  {Jungwirth}}, \ and\ \bibinfo {author} {\bibfnamefont {Jakub}\ \bibnamefont
  {\ifmmode~\check{Z}\else \v{Z}\fi{}elezn\'y}},\ }\bibfield  {title} {\enquote
  {\bibinfo {title} {Efficient electrical spin splitter based on
  nonrelativistic collinear antiferromagnetism},}\ }\href {\doibase
  10.1103/PhysRevLett.126.127701} {\bibfield  {journal} {\bibinfo  {journal}
  {Phys. Rev. Lett.}\ }\textbf {\bibinfo {volume} {126}},\ \bibinfo {pages}
  {127701} (\bibinfo {year} {2021})}\BibitemShut {NoStop}%
\bibitem [{\citenamefont {Liao}\ \emph {et~al.}(2024)\citenamefont {Liao},
  \citenamefont {Wang}, \citenamefont {Tien}, \citenamefont {Huang},\ and\
  \citenamefont {Qu}}]{Liao}%
  \BibitemOpen
  \bibfield  {author} {\bibinfo {author} {\bibfnamefont {Ching-Te}\
  \bibnamefont {Liao}}, \bibinfo {author} {\bibfnamefont {Yu-Chun}\
  \bibnamefont {Wang}}, \bibinfo {author} {\bibfnamefont {Yu-Cheng}\
  \bibnamefont {Tien}}, \bibinfo {author} {\bibfnamefont {Ssu-Yen}\
  \bibnamefont {Huang}}, \ and\ \bibinfo {author} {\bibfnamefont {Danru}\
  \bibnamefont {Qu}},\ }\bibfield  {title} {\enquote {\bibinfo {title}
  {Separation of inverse altermagnetic spin-splitting effect from inverse spin
  hall effect in ${\mathrm{ruo}}_{2}$},}\ }\href {\doibase
  10.1103/PhysRevLett.133.056701} {\bibfield  {journal} {\bibinfo  {journal}
  {Phys. Rev. Lett.}\ }\textbf {\bibinfo {volume} {133}},\ \bibinfo {pages}
  {056701} (\bibinfo {year} {2024})}\BibitemShut {NoStop}%
\bibitem [{\citenamefont {Yuan}\ \emph {et~al.}(2020)\citenamefont {Yuan},
  \citenamefont {Wang}, \citenamefont {Luo}, \citenamefont {Rashba},\ and\
  \citenamefont {Zunger}}]{Yuan_2020}%
  \BibitemOpen
  \bibfield  {author} {\bibinfo {author} {\bibfnamefont {Lin-Ding}\
  \bibnamefont {Yuan}}, \bibinfo {author} {\bibfnamefont {Zhi}\ \bibnamefont
  {Wang}}, \bibinfo {author} {\bibfnamefont {Jun-Wei}\ \bibnamefont {Luo}},
  \bibinfo {author} {\bibfnamefont {Emmanuel~I.}\ \bibnamefont {Rashba}}, \
  and\ \bibinfo {author} {\bibfnamefont {Alex}\ \bibnamefont {Zunger}},\
  }\bibfield  {title} {\enquote {\bibinfo {title} {Giant momentum-dependent
  spin splitting in centrosymmetric low-$z$ antiferromagnets},}\ }\href
  {\doibase 10.1103/PhysRevB.102.014422} {\bibfield  {journal} {\bibinfo
  {journal} {Phys. Rev. B}\ }\textbf {\bibinfo {volume} {102}},\ \bibinfo
  {pages} {014422} (\bibinfo {year} {2020})}\BibitemShut {NoStop}%
\bibitem [{\citenamefont {Giil}\ \emph {et~al.}(2024)\citenamefont {Giil},
  \citenamefont {Brekke}, \citenamefont {Linder},\ and\ \citenamefont
  {Brataas}}]{Giil_2024}%
  \BibitemOpen
  \bibfield  {author} {\bibinfo {author} {\bibfnamefont {Hans~Gl\o{}ckner}\
  \bibnamefont {Giil}}, \bibinfo {author} {\bibfnamefont {Bj\o{}rnulf}\
  \bibnamefont {Brekke}}, \bibinfo {author} {\bibfnamefont {Jacob}\
  \bibnamefont {Linder}}, \ and\ \bibinfo {author} {\bibfnamefont {Arne}\
  \bibnamefont {Brataas}},\ }\bibfield  {title} {\enquote {\bibinfo {title}
  {Quasiclassical theory of superconducting spin-splitter effects and
  spin-filtering via altermagnets},}\ }\href {\doibase
  10.1103/PhysRevB.110.L140506} {\bibfield  {journal} {\bibinfo  {journal}
  {Phys. Rev. B}\ }\textbf {\bibinfo {volume} {110}},\ \bibinfo {pages}
  {L140506} (\bibinfo {year} {2024})}\BibitemShut {NoStop}%
\bibitem [{\citenamefont {Kryder}\ \emph {et~al.}(2008)\citenamefont {Kryder},
  \citenamefont {Gage}, \citenamefont {McDaniel}, \citenamefont {Challener},
  \citenamefont {Rottmayer}, \citenamefont {Ju}, \citenamefont {Hsia},\ and\
  \citenamefont {Erden}}]{Kryder_2008}%
  \BibitemOpen
  \bibfield  {author} {\bibinfo {author} {\bibfnamefont {Mark~H.}\ \bibnamefont
  {Kryder}}, \bibinfo {author} {\bibfnamefont {Edward~C.}\ \bibnamefont
  {Gage}}, \bibinfo {author} {\bibfnamefont {Terry~W.}\ \bibnamefont
  {McDaniel}}, \bibinfo {author} {\bibfnamefont {William~A.}\ \bibnamefont
  {Challener}}, \bibinfo {author} {\bibfnamefont {Robert~E.}\ \bibnamefont
  {Rottmayer}}, \bibinfo {author} {\bibfnamefont {Ganping}\ \bibnamefont {Ju}},
  \bibinfo {author} {\bibfnamefont {Yiao-Tee}\ \bibnamefont {Hsia}}, \ and\
  \bibinfo {author} {\bibfnamefont {M.~Fatih}\ \bibnamefont {Erden}},\
  }\bibfield  {title} {\enquote {\bibinfo {title} {Heat assisted magnetic
  recording},}\ }\href {\doibase 10.1109/JPROC.2008.2004315} {\bibfield
  {journal} {\bibinfo  {journal} {Proceedings of the IEEE}\ }\textbf {\bibinfo
  {volume} {96}},\ \bibinfo {pages} {1810--1835} (\bibinfo {year}
  {2008})}\BibitemShut {NoStop}%
\bibitem [{\citenamefont {Sinova}\ \emph {et~al.}(2015)\citenamefont {Sinova},
  \citenamefont {Valenzuela}, \citenamefont {Wunderlich}, \citenamefont
  {Back},\ and\ \citenamefont {Jungwirth}}]{Sinova_2015}%
  \BibitemOpen
  \bibfield  {author} {\bibinfo {author} {\bibfnamefont {Jairo}\ \bibnamefont
  {Sinova}}, \bibinfo {author} {\bibfnamefont {Sergio~O.}\ \bibnamefont
  {Valenzuela}}, \bibinfo {author} {\bibfnamefont {J.}~\bibnamefont
  {Wunderlich}}, \bibinfo {author} {\bibfnamefont {C.~H.}\ \bibnamefont
  {Back}}, \ and\ \bibinfo {author} {\bibfnamefont {T.}~\bibnamefont
  {Jungwirth}},\ }\bibfield  {title} {\enquote {\bibinfo {title} {Spin hall
  effects},}\ }\href {\doibase 10.1103/RevModPhys.87.1213} {\bibfield
  {journal} {\bibinfo  {journal} {Rev. Mod. Phys.}\ }\textbf {\bibinfo {volume}
  {87}},\ \bibinfo {pages} {1213--1260} (\bibinfo {year} {2015})}\BibitemShut
  {NoStop}%
\bibitem [{\citenamefont {Liu}\ \emph {et~al.}(2011)\citenamefont {Liu},
  \citenamefont {Moriyama}, \citenamefont {Ralph},\ and\ \citenamefont
  {Buhrman}}]{Liu_2011}%
  \BibitemOpen
  \bibfield  {author} {\bibinfo {author} {\bibfnamefont {Luqiao}\ \bibnamefont
  {Liu}}, \bibinfo {author} {\bibfnamefont {Takahiro}\ \bibnamefont
  {Moriyama}}, \bibinfo {author} {\bibfnamefont {D.~C.}\ \bibnamefont {Ralph}},
  \ and\ \bibinfo {author} {\bibfnamefont {R.~A.}\ \bibnamefont {Buhrman}},\
  }\bibfield  {title} {\enquote {\bibinfo {title} {Spin-torque ferromagnetic
  resonance induced by the spin hall effect},}\ }\href {\doibase
  10.1103/PhysRevLett.106.036601} {\bibfield  {journal} {\bibinfo  {journal}
  {Phys. Rev. Lett.}\ }\textbf {\bibinfo {volume} {106}},\ \bibinfo {pages}
  {036601} (\bibinfo {year} {2011})}\BibitemShut {NoStop}%
\bibitem [{\citenamefont {Torques}(2005)}]{Torques_2005}%
  \BibitemOpen
  \bibfield  {author} {\bibinfo {author} {\bibfnamefont {Spin-Transfer}\
  \bibnamefont {Torques}},\ }\bibfield  {title} {\enquote {\bibinfo {title}
  {Time-domain measurements of nanomagnet dynamics driven by},}\ }\href
  {https://www.science.org/doi/10.1126/science.1105722} {\bibfield  {journal}
  {\bibinfo  {journal} {science}\ }\textbf {\bibinfo {volume} {1105722}},\
  \bibinfo {pages} {307} (\bibinfo {year} {2005})}\BibitemShut {NoStop}%
\bibitem [{\citenamefont {Lopusnik}\ \emph {et~al.}(2003)\citenamefont
  {Lopusnik}, \citenamefont {Nibarger}, \citenamefont {Silva},\ and\
  \citenamefont {Celinski}}]{Lopusnik_2003}%
  \BibitemOpen
  \bibfield  {author} {\bibinfo {author} {\bibfnamefont {R.}~\bibnamefont
  {Lopusnik}}, \bibinfo {author} {\bibfnamefont {J.~P.}\ \bibnamefont
  {Nibarger}}, \bibinfo {author} {\bibfnamefont {T.~J.}\ \bibnamefont {Silva}},
  \ and\ \bibinfo {author} {\bibfnamefont {Z.}~\bibnamefont {Celinski}},\
  }\bibfield  {title} {\enquote {\bibinfo {title} {Different dynamic and static
  magnetic anisotropy in thin permalloy™ films},}\ }\href {\doibase
  10.1063/1.1587255} {\bibfield  {journal} {\bibinfo  {journal} {Applied
  Physics Letters}\ }\textbf {\bibinfo {volume} {83}},\ \bibinfo {pages}
  {96--98} (\bibinfo {year} {2003})}\BibitemShut {NoStop}%
\bibitem [{\citenamefont {\textit{et al.}}(2019)}]{Peng_2019}%
  \BibitemOpen
  \bibfield  {author} {\bibinfo {author} {\bibfnamefont {W.~L.~Peng}\
  \bibnamefont {\textit{et al.}}},\ }\bibfield  {title} {\enquote {\bibinfo
  {title} {Tunable damping-like and field-like spin-orbit-torque in pt/co/hfo2
  films via interfacial charge transfer},}\ }\href {\doibase 10.1063/1.5123018}
  {\bibfield  {journal} {\bibinfo  {journal} {Applied Physics Letters}\
  }\textbf {\bibinfo {volume} {115}},\ \bibinfo {pages} {172403} (\bibinfo
  {year} {2019})}\BibitemShut {NoStop}%
\bibitem [{\citenamefont {Husain}\ and\ \citenamefont {\textit{et
  al.}}(2020)}]{Husain_2020}%
  \BibitemOpen
  \bibfield  {author} {\bibinfo {author} {\bibfnamefont {Sajid}\ \bibnamefont
  {Husain}}\ and\ \bibinfo {author} {\bibfnamefont {Xin~Chen}\ \bibnamefont
  {\textit{et al.}}},\ }\bibfield  {title} {\enquote {\bibinfo {title} {Large
  damping-like spin–orbit torque in a 2d conductive 1t-tas2 monolayer},}\
  }\href {\doibase 10.1021/acs.nanolett.0c01955} {\bibfield  {journal}
  {\bibinfo  {journal} {Nano Letters}\ }\textbf {\bibinfo {volume} {20}},\
  \bibinfo {pages} {6372--6380} (\bibinfo {year} {2020})}\BibitemShut {NoStop}%
\bibitem [{\citenamefont {Gong}\ and\ \citenamefont
  {Zhang}(2019)}]{Cheng_2019}%
  \BibitemOpen
  \bibfield  {author} {\bibinfo {author} {\bibfnamefont {Cheng}\ \bibnamefont
  {Gong}}\ and\ \bibinfo {author} {\bibfnamefont {Xiang}\ \bibnamefont
  {Zhang}},\ }\bibfield  {title} {\enquote {\bibinfo {title} {Two-dimensional
  magnetic crystals and emergent heterostructure devices},}\ }\href {\doibase
  10.1126/science.aav4450} {\bibfield  {journal} {\bibinfo  {journal}
  {Science}\ }\textbf {\bibinfo {volume} {363}},\ \bibinfo {pages} {eaav4450}
  (\bibinfo {year} {2019})}\BibitemShut {NoStop}%
\bibitem [{\citenamefont {\textit{et al.}}(2024{\natexlab{b}})}]{Hariki_2024}%
  \BibitemOpen
  \bibfield  {author} {\bibinfo {author} {\bibfnamefont {A.~Hariki}\
  \bibnamefont {\textit{et al.}}},\ }\bibfield  {title} {\enquote {\bibinfo
  {title} {X-ray magnetic circular dichroism in altermagnetic
  $\ensuremath{\alpha}$-mnte},}\ }\href {\doibase
  10.1103/PhysRevLett.132.176701} {\bibfield  {journal} {\bibinfo  {journal}
  {Phys. Rev. Lett.}\ }\textbf {\bibinfo {volume} {132}},\ \bibinfo {pages}
  {176701} (\bibinfo {year} {2024}{\natexlab{b}})}\BibitemShut {NoStop}%
\bibitem [{\citenamefont {Sheoran}\ and\ \citenamefont
  {Bhattacharya}(2024)}]{Sheoran_2024}%
  \BibitemOpen
  \bibfield  {author} {\bibinfo {author} {\bibfnamefont {Sajjan}\ \bibnamefont
  {Sheoran}}\ and\ \bibinfo {author} {\bibfnamefont {Saswata}\ \bibnamefont
  {Bhattacharya}},\ }\bibfield  {title} {\enquote {\bibinfo {title}
  {Nonrelativistic spin splittings and altermagnetism in twisted bilayers of
  centrosymmetric antiferromagnets},}\ }\href {\doibase
  10.1103/PhysRevMaterials.8.L051401} {\bibfield  {journal} {\bibinfo
  {journal} {Phys. Rev. Mater.}\ }\textbf {\bibinfo {volume} {8}},\ \bibinfo
  {pages} {L051401} (\bibinfo {year} {2024})}\BibitemShut {NoStop}%
\bibitem [{\citenamefont {Reimers}\ \emph {et~al.}(2024)\citenamefont
  {Reimers}, \citenamefont {Odenbreit}, \citenamefont {{\v{S}}mejkal},
  \citenamefont {Strocov}, \citenamefont {Constantinou}, \citenamefont
  {Hellenes}, \citenamefont {Jaeschke~Ubiergo}, \citenamefont {Campos},
  \citenamefont {Bharadwaj}, \citenamefont {Chakraborty} \emph
  {et~al.}}]{Reimers_2024}%
  \BibitemOpen
  \bibfield  {author} {\bibinfo {author} {\bibfnamefont {Sonka}\ \bibnamefont
  {Reimers}}, \bibinfo {author} {\bibfnamefont {Lukas}\ \bibnamefont
  {Odenbreit}}, \bibinfo {author} {\bibfnamefont {Libor}\ \bibnamefont
  {{\v{S}}mejkal}}, \bibinfo {author} {\bibfnamefont {Vladimir~N}\ \bibnamefont
  {Strocov}}, \bibinfo {author} {\bibfnamefont {Procopios}\ \bibnamefont
  {Constantinou}}, \bibinfo {author} {\bibfnamefont {Anna~B}\ \bibnamefont
  {Hellenes}}, \bibinfo {author} {\bibfnamefont {Rodrigo}\ \bibnamefont
  {Jaeschke~Ubiergo}}, \bibinfo {author} {\bibfnamefont {Warlley~H}\
  \bibnamefont {Campos}}, \bibinfo {author} {\bibfnamefont {Venkata~K}\
  \bibnamefont {Bharadwaj}}, \bibinfo {author} {\bibfnamefont {Atasi}\
  \bibnamefont {Chakraborty}},  \emph {et~al.},\ }\bibfield  {title} {\enquote
  {\bibinfo {title} {Direct observation of altermagnetic band splitting in crsb
  thin films},}\ }\href {https://www.nature.com/articles/s41467-024-46476-5}
  {\bibfield  {journal} {\bibinfo  {journal} {Nature Communications}\ }\textbf
  {\bibinfo {volume} {15}},\ \bibinfo {pages} {2116} (\bibinfo {year}
  {2024})}\BibitemShut {NoStop}%
\bibitem [{\citenamefont {Liu}\ \emph {et~al.}(2024{\natexlab{b}})\citenamefont
  {Liu}, \citenamefont {Yu},\ and\ \citenamefont {Liu}}]{Liu_2024_}%
  \BibitemOpen
  \bibfield  {author} {\bibinfo {author} {\bibfnamefont {Yichen}\ \bibnamefont
  {Liu}}, \bibinfo {author} {\bibfnamefont {Junxi}\ \bibnamefont {Yu}}, \ and\
  \bibinfo {author} {\bibfnamefont {Cheng-Cheng}\ \bibnamefont {Liu}},\
  }\bibfield  {title} {\enquote {\bibinfo {title} {Twisted magnetic van der
  waals bilayers: An ideal platform for altermagnetism},}\ }\href {\doibase
  10.1103/PhysRevLett.133.206702} {\bibfield  {journal} {\bibinfo  {journal}
  {Phys. Rev. Lett.}\ }\textbf {\bibinfo {volume} {133}},\ \bibinfo {pages}
  {206702} (\bibinfo {year} {2024}{\natexlab{b}})}\BibitemShut {NoStop}%
\bibitem [{\citenamefont {Pan}\ \emph {et~al.}(2024)\citenamefont {Pan},
  \citenamefont {Zhou}, \citenamefont {Lyu}, \citenamefont {Xiao},
  \citenamefont {Yang},\ and\ \citenamefont {Sun}}]{Pan_2024}%
  \BibitemOpen
  \bibfield  {author} {\bibinfo {author} {\bibfnamefont {Baoru}\ \bibnamefont
  {Pan}}, \bibinfo {author} {\bibfnamefont {Pan}\ \bibnamefont {Zhou}},
  \bibinfo {author} {\bibfnamefont {Pengbo}\ \bibnamefont {Lyu}}, \bibinfo
  {author} {\bibfnamefont {Huaping}\ \bibnamefont {Xiao}}, \bibinfo {author}
  {\bibfnamefont {Xuejuan}\ \bibnamefont {Yang}}, \ and\ \bibinfo {author}
  {\bibfnamefont {Lizhong}\ \bibnamefont {Sun}},\ }\bibfield  {title} {\enquote
  {\bibinfo {title} {General stacking theory for altermagnetism in bilayer
  systems},}\ }\href {\doibase 10.1103/PhysRevLett.133.166701} {\bibfield
  {journal} {\bibinfo  {journal} {Phys. Rev. Lett.}\ }\textbf {\bibinfo
  {volume} {133}},\ \bibinfo {pages} {166701} (\bibinfo {year}
  {2024})}\BibitemShut {NoStop}%
\bibitem [{\citenamefont {Haldane}(1988)}]{Haldane_1998}%
  \BibitemOpen
  \bibfield  {author} {\bibinfo {author} {\bibfnamefont {F.~D.~M.}\
  \bibnamefont {Haldane}},\ }\bibfield  {title} {\enquote {\bibinfo {title}
  {Model for a quantum hall effect without landau levels: Condensed-matter
  realization of the "parity anomaly"},}\ }\href {\doibase
  10.1103/PhysRevLett.61.2015} {\bibfield  {journal} {\bibinfo  {journal}
  {Phys. Rev. Lett.}\ }\textbf {\bibinfo {volume} {61}},\ \bibinfo {pages}
  {2015--2018} (\bibinfo {year} {1988})}\BibitemShut {NoStop}%
\end{thebibliography}%

\end{document}